\documentclass[11pt]{article}

\usepackage[square,authoryear]{natbib}
\usepackage{marsden_article}
\usepackage{wrapfig}

\usepackage[margin=1cm, font=small,format=plain,labelfont=bf,up,textfont=up]{caption}
\usepackage{subfig}
\usepackage{graphicx}

\renewcommand{\Im}{\operatorname{Im}}
\renewcommand{\Re}{\operatorname{Re}}

\title{A novel formulation of point vortex dynamics \\
on the sphere: geometrical and numerical aspects}
\author{Joris Vankerschaver\thanks{Current address: Department of Mathematics, Imperial College London, London SW7 2AZ, United Kingdom. Email: \url{joris.vankerschaver@gmail.com}.}, Melvin Leok\thanks{Email: \url{mleok@math.ucsd.edu}.}  \\
{\small Department of Mathematics} \\
{\small University of California, San Diego} \\
{\small 9500 Gilman Drive} \\
{\small La Jolla, CA 92093-0112}
}

\begin{document}
\maketitle

\begin{abstract}
	In this paper, we present a novel Lagrangian formulation of the equations of motion for point vortices on the unit 2-sphere.  We show first that no linear Lagrangian formulation exists directly on the 2-sphere but that a Lagrangian may be constructed by pulling back the dynamics to the 3-sphere by means of the Hopf fibration.  We then use the isomorphism of the 3-sphere with the Lie group $SU(2)$ to derive a variational Lie group integrator for point vortices which is symplectic, second-order, and preserves the unit-length constraint.  At the end of the paper, we compare our integrator with classical fourth-order Runge--Kutta, the second-order midpoint method, and a standard Lie group Munthe-Kaas method.
\end{abstract}

\section{Introduction}

Point vortices are point-like singularities in the vorticity field of an ideal fluid.  First described by \cite{He1858}, they form a finite-dimensional singular solution of the Euler equations and are now a classical subject in hydrodynamics, see among others \cite{La1945, Mi1968, Sa1992, Ne2001}.  

The interest in point vortices is two-fold.   On the one hand, paraphrasing \cite{Ar2007}, the description of point vortices forms a veritable playground for classical mathematics and gives rise to interesting phenomena from dynamical systems, such as periodic motions (\cite{SoTo2002, BoMaKi2004}), (relative) equilibria (\cite{PoDr1993, KiNe1998, Ar2011}), and chaotic advection and topological chaos in fluids (\cite{BoStAr2003}).  On the numerical front, on the other hand, desingularizations of the point vortex equations, such as the classical vortex blob method of \cite{Ch1973} form the basis for important classes of particle methods for the Euler and Navier-Stokes equations.  The idea is that the vorticity field of an arbitrary fluid can be approximated by a number of vortex blobs whose motion is then followed in time.  Strong analytical estimates exist that link the behavior of the vortex blobs with the solution of the Euler equations that they approximate (\cite{MaBe2002}).

On the sphere, the dynamics of point vortices was first described by \cite{Bo1977} after a model by Gromeka (see \cite{Ne2001} for an historical overview) and is in some sense a generalization of the planar case (see also \cite{KiOk1987} and \cite{PoDr1993}).  The relevance of point vortices of the sphere lies in the fact that they provide a first approximation of the behavior of certain geophysical flows for which the curvature of the earth is important, and which persist over long periods of time.   The mathematical description of point vortices on the sphere is an area of active research: fixed and relative equilibria of the three-vortex problem were described in \cite{PoDr1993, KiNe1998} (see also \cite{PeMa1998}), while more general equilibria were described in \cite{LiMoRo2001,ChKaNe2009, NeSa2011}.  Conditions for the collapse of point vortex configurations on the sphere were established in \cite{KiNe1998} and \cite{Sa2008}.

Most of the research on point vortices on the sphere has focused on the existence of analytical solutions such as relative equilibria for few point vortices, but comparatively little is known about the behavior of arbitrary configurations of vortices.  One of the contributions of this paper is to construct a geometric numerical integrator which is second-order accurate, preserves the geometry of the sphere, and is symplectic.  As symplectic integrators are known to capture the long-term behavior of a Hamiltonian system better than classical integrators (see \cite{PuSa1991} for an application of symplectic integrators to point vortex dynamics in the plane, and \cite{MaWe2001, HaLuWa2002} for a general overview of variational integration techniques), we expect our geometric integrator to give insight into the behavior of non-equilibrium vortex configurations, even over long integration times.

\subsection{Aims and contributions of this paper}

The contributions of this paper are two-fold.  In the first part of this paper, we construct a Lagrangian description for point vortices on the sphere in terms of pairs of complex numbers.  We first review the Lagrangian description for point vortices in the plane (see e.g. \cite{Ch1978, Ne2001, RoMa2002}) and then show via a simple topological argument that no (linear) Lagrangian exists for the dynamics of point vortices on the two-dimensional sphere $\mathbb{S}^2$.

We then use the \emph{Hopf fibration}, a distinguished submersion from the three-sphere $\mathbb{S}^3$ to the two-sphere $\mathbb{S}^2$, to pull back the Hamiltonian description to $\mathbb{S}^3$, where the topological obstruction for the existence of a linear Lagrangian vanishes.  We explicitly construct this Lagrangian and we show that the equations of motion give rise to a (finite-dimensional) \emph{non-linear Schr\"odinger equation} on $\mathbb{S}^3$ with gauge freedom.  These equations bear a remarkable similarity to the equations of motion for point vortices in the complex plane, only now the location of each point vortex is specified by a pair of complex numbers (or equivalently, a (unit) quaternion) instead of a single one. 

In the second part of the paper, we design a variational numerical integrator for point vortices on the sphere using the linear Lagrangian on $\mathbb{S}^3$.  We use the identification between the 3-sphere $\mathbb{S}^3$ and the Lie group $SU(2)$ of special unitary 2-by-2 matrices to write the update equation for the integrator as a fixed-point equation in the Lie algebra $\mathfrak{su}(2)$, and we show how the discrete equations of motion are symplectic, self-adjoint, second-order, and preserve the unit-length constraint in $\mathbb{S}^3$.  At the end of the paper, we compare our integrator to the classical 4th order Runge--Kutta method, as well as to a number of geometric integration methods.  We show that the geometric integrators, and in particular the Hopf variational integrator, outperform Runge--Kutta in the medium run, even though they are only second-order accurate.

\subsection{Background and historical overview}

\paragraph{Linear Lagrangian formulation for planar vortices.}

We review here the Lagrangian and Hamiltonian descriptions of point vortices in the plane.  The Hamiltonian description of point vortices on the sphere will be reviewed in Section~\ref{sec:hamiltonian}.

For point vortices in the plane, the equations of motion are given in complex form by 
\begin{equation} \label{plane_equations}
	\dot{z}_\alpha = -2 \mathrm{i} \frac{\partial H}{\partial z_\alpha^\ast}.
\end{equation}
Here, the $z_\alpha$ ($\alpha = 1, \ldots, N$) represent the locations in the complex plane of each of the vortices, and $\Gamma_\alpha$ is a real parameter which specifies the circulation around each vortex.  The Hamiltonian function is given by
\begin{equation} \label{plane_hamiltonian}
	H(z_1, \ldots, z_N) = -\frac{1}{4\pi} \sum_{\alpha < \beta} \Gamma_\alpha \Gamma_\beta \log \left| z_\alpha - z_\beta \right|^2.
\end{equation}

These equations can be derived from a Lagrangian which is \emph{linear} in the velocities (see \cite{Ch1978}) and is given by 
\begin{equation} \label{plane_lagrangian}
	L = 
	\frac{1}{2\mathrm{i}} \sum_{\alpha=1}^N \Gamma_\alpha 
		( z_\alpha^\ast \dot{z}_\alpha 
		 - z_\alpha \dot{z}^\ast_\alpha) - H(z_1, \ldots, z_N). 
\end{equation} 
For future reference, we point out that the linear part of the Lagrangian can be written as $\sum \Gamma_\alpha \theta(z_\alpha, \dot{z}_\alpha)$, where $\theta$ is the one-form given by 
\begin{equation} \label{complex_one_form}
	\theta = \frac{1}{2} \Im(z^\ast  dz).
\end{equation}
The exterior derivative of $\theta$ is nothing but the area form on the complex plane: 
\begin{equation} \label{area_form_plane}
	\mathbf{d} \theta = \frac{1}{2} \Im(dz^\ast \wedge dz) = dx \wedge dy.
\end{equation}
It can be shown that the flow of the point vortex equations \eqref{plane_equations} preserves a weighted sum of such area forms, given by  
\[
	\sum_{\alpha = 1}^N \Gamma_\alpha \mathbf{d} \theta_\alpha = \sum_{\alpha=1} \Gamma_\alpha dx_\alpha \wedge dy_\alpha,
\]
where $\mathbf{d} \theta_\alpha$ refers to the area form \eqref{area_form_plane} expressed in the coordinates of the $\alpha$th vortex.  This is an example of a symplectic form on the phase space $\mathbb{C}^N$.

The advantage of having a Lagrangian description for the dynamics of point vortices is that the standard results for the construction of Lagrangian variational integrators (see \cite{MaWe2001} for an overview) can now be applied.  This is the key observation of \cite{RoMa2002}, who constructed a class of second-order variational integrators by discretizing the Lagrangian \eqref{plane_lagrangian} using centered finite differences. 

Before turning to the case of point vortices on the sphere, we point out that many non-canonical Hamiltonian systems can be rephrased as Euler--Lagrange equations that come from a Lagrangian which is linear in the velocities.  This observation was made by \cite{Bi1966} in his study of Pfaffian systems and was used in \cite{FaJa1988} as a starting point for the description of Hamiltonian systems with constraints.  Linear Lagrangians also appear in the description of the non-linear Schr\"odinger equation and the KdV equation.

\paragraph{The dynamics of point vortices on the sphere.}

The equations of motion for $N$ point vortices with strengths $\Gamma_i$, $i = 1, \ldots, N$ on the unit sphere $\mathbb{S}^2$ can be written as follows (see  \cite{Ne2001}).  If we denote the position vector of the $i$th vortex by $\mathbf{x}_i$ (so that $\left\Vert \mathbf{x}_i \right\Vert = 1$), the point vortex equations can be written in Euclidian form as
\begin{equation} \label{vortexequations}
	 \dot{\mathbf{x}}_k = \frac{1}{4 \pi} \sum_{j \ne k} \Gamma_j  \frac{ \mathbf{x}_j \times  \mathbf{x}_k}{1 + \sigma^2 -  \mathbf{x}_k \cdot  \mathbf{x}_j},
\end{equation}
where $\sigma$ is a small regularization parameter which is added to ensure that the limit of the right-hand side exists when $\mathbf{x}_k$ tends to $\mathbf{x}_j$.  Note that the equations \eqref{vortexequations} conserve the \emph{vortex moment}, defined as 
\[
	\mathbf{M} = \sum_{i = 1}^N \Gamma_i \mathbf{x}_i.
\]
We also point out that due to topological reasons, the total vorticity on the sphere must be zero (see \cite{Ne2001, BoKo2008}). If the sum of the strengths of the point vortices is not zero, $\sum_{i=1}^N \Gamma_i \ne 0$, then the full vorticity field on the sphere will have counterrotating point vortices or patches of vorticity to balance the effect of the point vortices.

\paragraph{Non-existence of a linear Lagrangian for vortices on the sphere.}

In Section~\ref{sec:hamiltonian}, we will review the Hamiltonian formulation for the point vortex equations \eqref{vortexequations}.  We now discuss the Lagrangian formulation, and in particular we argue that \emph{no linear Lagrangian exists for the dynamics of point vortices on $\mathbb{S}^2$}. This can be seen by the fact that a linear Lagrangian on $(\mathbb{S}^2)^N$ would necessarily have to be of the form $L = \mathcal{A}_\Gamma - H$, with $\mathcal{A}_\Gamma$ a one-form on $(\mathbb{S}^2)^N$. The symplectic form preserved by the flow of the Euler-Lagrange equations would then be $\mathbf{d} \mathcal{A}_\Gamma$, which is by definition  exact. However, a simple topological argument can be used to show that on $(\mathbb{S}^2)^N$, or on any compact manifold, any symplectic form must be non-exact. We reproduce this argument for the case of point vortices below; see \cite{McSa1998} for the general case.

For point vortices on the sphere, the phase space is the product $(\mathbb{S}^2)^N$ of $N$ copies of the unit sphere $\mathbb{S}^2$, equipped with a symplectic form $\mathcal{B}_\Gamma$ which is a weighted sum of the area forms on the individual spheres:  
\[
	\mathcal{B}_\Gamma = \sum_{i = 1}^N \Gamma_i \Omega_i,
\] 
where $\Omega_i$ is the area form on the $i$th copy of $\mathbb{S}^2$.

As $(\mathbb{S}^2)^N$ is compact, this form cannot be exact. The argument to see this is as follows (see \cite{McSa1998}): integrate the symplectic volume form 
\[
	\mathcal{B}_\Gamma^N 
	:= \frac{1}{N!} \mathcal{B}_\Gamma \wedge \cdots \wedge \mathcal{B}_\Gamma 
		= \left( \Pi_{i = 1}^N \Gamma_i \right) 
			\Omega_1 \wedge \cdots \wedge \Omega_N
\]
over the entire phase space to get 
\[
	\int_{(\mathbb{S}^2)^N} \mathcal{B}_\Gamma^N = 
		(4 \pi)^N \left( \Pi_{i = 1}^N \Gamma_i \right) \ne 0.
\]
On the other hand, if the symplectic form $\mathcal{B}_\Gamma$ were exact, $\mathcal{B}_\Gamma = \mathbf{d} \mathcal{A}_\Gamma$, then $\mathcal{B}_\Gamma^N$ would be exact too, since in this case $\mathcal{B}_\Gamma^N = 1/N! \, \mathbf{d}( \mathcal{A}_\Gamma \wedge \mathcal{B}_\Gamma \wedge \cdots \wedge \mathcal{B}_\Gamma)$. In this case, integrating over $(\mathbb{S}^2)^N$ would result in zero symplectic volume because of Stokes' theorem, a contradiction. 

One way out is as follows. Below, we will see that the area symplectic form $\Omega$ on $\mathbb{S}^2$ can be pulled back to an exact two-form on the three-sphere $\mathbb{S}^3$.  This will allow us to construct a linear Lagrangian for vortical structures on $(\mathbb{S}^3)^N$, and the solutions of the Euler-Lagrange equations for this Lagrangian will be seen to project down onto solutions of the point vortex equations on $(\mathbb{S}^2)^N$. By discretizing the Lagrangian variational principle on $(\mathbb{S}^3)^N$ (using the techniques from \cite{MaWe2001} and \cite{LeLeMc2007}),  we will then be able to construct a variational integrator for point vortices which is automatically symplectic, second-order, and unit-length preserving.

\paragraph{Other approaches to the numerical integration of point vortices.}

The use of symplectic methods in vortex dynamics was pioneered by
\cite{PuSa1991}, who used a fourth-order symplectic Runge-Kutta scheme
to integrate the equations of motion for four vortices in the
plane. It is not clear, however, how to extend their method to the
case of vortices on the sphere.

Hamiltonian variational principles have been developed by
\cite{Oh1997} in the context of Floer homology and by \cite{No1982}
for Morse theory (see also \cite{CeMa1987}).  On the numerical front,
geometrical numerical integration of Hamiltonian systems was described
in \cite{Brown2006}, \cite{MaRo2010} and \cite{LeZh2011}, but all of
these references assume that the underlying symplectic manifold is
exact.  For non-exact symplectic forms (e.g. the case of point
vortices on the sphere) it is as of yet not clear how to discretize
the Hamiltonian variational principle so that the resulting numerical
algorithms share some of the properties of the continuous system (such
as symplecticity and momentum preservation).

We do remark that \cite{MaRo2010} perform a similar pullback as in
this paper, but using the Lie algebra of the rotation group $SO(3)$
instead of the special unitary group $SU(2)$, in order to make the
dynamics of point vortices on the sphere amenable to geometric
integration.

\section{The Hopf fibration} \label{sec:hopf}

In this section, we introduce our notation and review some aspects of the geometry of the spheres $\mathbb{S}^2, \mathbb{S}^3$ and the Hopf fibration.  This material is standard and can be found in any geometric physics textbook, for instance \cite{Fr2004}.  More information about the Hopf fibration and its role in physics and geometry can be found in \cite{Mo2002, Ur2003, Le1948} and the references therein.

\paragraph{Notation.}

We will denote vectors in $\mathbb{C}^2$ and their Hermitian conjugates by 
\[
	\varphi := \begin{bmatrix} z \\ u \end{bmatrix}, 
		\quad \text{and} \quad 
	{\varphi}^\dagger := \begin{bmatrix} {z}^\ast, & {u}^\ast \end{bmatrix},
\]
where $z^\ast$ is the complex conjugate of $z \in \mathbb{C}$.  The Hermitian conjugate of a complex matrix $A$ will be denoted by $A^\dagger$.

Lowercase Roman letters $a, b, \dots$ will refer to the components $\varphi^a$ of a vector $\varphi$ in $\mathbb{C}^2$.  The Greek letters $\alpha, \beta, \dots$ will refer to the Cartesian components $x^\alpha$ of a vector $\mathbf{x} \in \mathbb{R}^3$.  The imaginary unit will be denoted by $\mathrm{i}$.

The Hermitian inner product on $\mathbb{C}^2$ is given by 
\[
	\left< \varphi_1, \varphi_2 \right> := \varphi_1^\dagger \varphi_2 = {z}^\ast_1 z_2 + {u}^\ast_1 u_2.
\]
Note that the Euclidian inner product on $\mathbb{C}^2$ can be expressed as 
\begin{equation} \label{euclidian_inner_product}
	\Re \left< \varphi_1, \varphi_2 \right> = \Re( {z}^\ast_1 z_2 + {u}^\ast_1 u_2).
\end{equation}

\paragraph{The geometry of $\mathbb{S}^2$.}  

We think of the two-sphere $\mathbb{S}^2$ as the set of all unit-length vectors $\mathbf{x}$ in $\mathbb{R}^3$.  The tangent plane $T_{\mathbf{x}} \mathbb{S}^2$ at an element $\mathbf{x} \in \mathbb{S}^2$ consists of all vectors, denoted by $\delta \mathbf{x} \in \mathbb{R}^3$, which are orthogonal to $\mathbf{x}$:
\[
	T_{\mathbf{x}} \mathbb{S}^2 = \{ \delta \mathbf{x} \in \mathbb{R}^3: \mathbf{x} \cdot \delta \mathbf{x} = 0 \}.
\]

In Cartesian coordinates, the area form $\Omega$ on $\mathbb{S}^2$ can be described as follows: $\Omega$ is the differential two-form given by 
\begin{equation} \label{area_form}
	\Omega(\mathbf{x})(\delta \mathbf{x}, \delta \mathbf{y}) 
	= \mathbf{x} \cdot 
		(\delta \mathbf{x} \times \delta \mathbf{y}) ,
\end{equation}
for all $\mathbf{x} \in \mathbb{S}^2$ and $\delta \mathbf{x}, \delta \mathbf{y} \in T_{\mathbf{x}} \mathbb{S}^2$. In spherical coordinates, $\Omega = \sin \theta d \theta \wedge d \phi$.  Note that $\Omega$ is not exact.

\paragraph{The geometry of $\mathbb{S}^3$ and the group $SU(2)$.}

We let $\mathbb{S}^3$ be the unit sphere in $\mathbb{C}^2$: 
\[
	\mathbb{S}^3 = \{ (z, u) \in \mathbb{C}^2: |z|^2 + |u|^2 = 1 \}.
\]
The tangent plane at an element $\varphi \in \mathbb{S}^3$ is given by the set of all vectors, denoted by $\delta \varphi \in \mathbb{C}^2$, which are orthogonal to $\varphi$:
\begin{equation} \label{tangent_space_s3}
	T_\varphi \mathbb{S}^3 := \{ \delta \varphi \in \mathbb{C}^2: \Re \left< \delta \varphi, \varphi \right> = 0 \},
\end{equation}
where we have expressed the orthogonality between $\varphi$ and $\delta \varphi$ using the inner product \eqref{euclidian_inner_product} in $\mathbb{C}^2$.

The unit sphere $\mathbb{S}^3$ can be embedded into the complex 2-by-2 matrices by means of the mapping  
\[
	 \begin{bmatrix} z \\ u \end{bmatrix} \in \mathbb{S}^3 \mapsto  
		\begin{bmatrix} 
			z & -u^\ast \\
			u & z^\ast 
		\end{bmatrix} \in M_2(\mathbb{C}),
\]
whose range is precisely the Lie group $SU(2)$ consisting of all Hermitian matrices ($U^\dagger = U$) with unit determinant ($\det U = 1$).  The Lie algebra of $SU(2)$ is the vector space $\mathfrak{su}(2)$, consisting of all 2-by-2 matrices $A$ which are anti-Hermitian ($A^\dagger = -A$) and traceless ($\mathrm{tr} A = 0$).  The identification of $\mathbb{S}^3$ with $SU(2)$ provides a convenient description for the tangent spaces \eqref{tangent_space_s3}: we have that $\delta \varphi \in T_\varphi \mathbb{S}^3$ if and only if there is a matrix $A \in \mathfrak{su}(2)$ such that 
\begin{equation} \label{su_variations}
	\delta \varphi = A \varphi.
\end{equation}
To see this, note that $A^\dagger = -A$ implies that $\left<\varphi, A\varphi\right>$ is purely imaginary, so that $\Re \left<\varphi, A\varphi\right> = 0$.

The Lie algebra $\mathfrak{su}(2)$ has dimension 3 and a convenient basis is given by the matrices $\tau_\alpha = \mathrm{i} \sigma_\alpha$, $\alpha = 1, 2, 3$, where the $\sigma_\alpha$ are the \emph{Pauli spin matrices}:
\begin{equation} \label{pauli}
	\sigma_1 = 
		\begin{bmatrix}
			0 & 1 \\
			1 & 0 
		\end{bmatrix}, 
	\quad
	\sigma_2 = 
		\begin{bmatrix}
			0 & -\mathrm{i}  \\
			\mathrm{i}  & 0 
		\end{bmatrix},
	\quad \text{and} \quad 
	\sigma_3 = 
		\begin{bmatrix}
			1 & 0 \\
			0 & -1
		\end{bmatrix}.
\end{equation}
Given a matrix $A \in \mathfrak{su}(2)$, we will denote its components in this basis by ${a}_\alpha$, $\alpha = 1, 2, 3$, and we put $\mathbf{a} := (a_1, a_2, a_3)$.  Explicitly, 
\begin{equation} \label{vectorrep}
	A = \mathbf{a} \cdot (\mathrm{i}\boldsymbol{\sigma}) = \sum_{\alpha = 1}^3 a_\alpha (\mathrm{i} \sigma_\alpha),
\end{equation}
where $\boldsymbol{\sigma}$ represents the vector $(\sigma^1, \sigma^2, \sigma^3)$.  We will refer to $\mathbf{a} \in \mathbb{R}^3$ as the \emph{vector representation} of the matrix $A \in \mathfrak{su}(2)$.

The Pauli matrices satisfy a number of useful identities:  the multiplication identity is
\begin{equation} \label{pauli_mult}
	\sigma_\alpha \sigma_\beta = \delta_{\alpha \beta} I + \mathrm{i} \sum_{\gamma = 1}^3 \epsilon_{\alpha\beta\gamma} \sigma_\gamma,
\end{equation}
for $\alpha, \beta = 1, 2, 3$,  where $I$ is the 2-by-2 unit matrix and $\epsilon_{\alpha\beta\gamma}$ the Levi-Civita permutation symbol.  Secondly, there is  the completeness property
\begin{equation} \label{pauli_comp}
	\sum_{\alpha = 1}^3 (\sigma_\alpha)_{ab} (\sigma_\alpha)_{cd} = 2 \delta_{ad} \delta_{bc} - \delta_{ab} \delta_{cd},
\end{equation}
for all $a, b, c, d = 1, 2$.
Proofs of these identities can be found in any standard textbook on quantum mechanics.

\paragraph{The Hopf fibration.} 

The group $ U(1) \cong \mathbb{S}^1$ of unit complex numbers acts on $\mathbb{S}^3$ by the diagonal action: $e^{i \theta} \cdot (z, u) = (e^{i\theta} z, e^{i\theta} u)$ for all $e^{i \theta} \in \mathbb{S}^1$ and $(z, u) \in \mathbb{S}^3$.  In terms of $SU(2)$-matrices, this action is described as 
\begin{equation} \label{circle_action}
	\begin{bmatrix} 
		z & -u^\ast \\
		u & z^\ast 
	\end{bmatrix} \cdot e^{\mathrm{i}\theta} 
	= \begin{bmatrix} 
			z & -u^\ast \\
			u & z^\ast 
		\end{bmatrix}
	\begin{bmatrix} 
			e^{\mathrm{i}\theta} & 0 \\
			0 & e^{-\mathrm{i}\theta}
		\end{bmatrix}.
\end{equation}

 The orbit space of this action, $\mathbb{S}^3/\mathbb{S}^1$, can be identified with the two-sphere $\mathbb{S}^2$.  Explicitly, there exists a surjective submersion $\pi: \mathbb{S}^3 \to \mathbb{S}^2$, called the \emph{Hopf fibration}, given by 
\begin{equation} \label{projection}
	\pi(z, u) = (2 \Re({z}^\ast u), 2 \Im({z}^\ast u), |z|^2 - |u|^2), 
\end{equation}
and the fibers of $\pi$ coincide with the orbits of the group $\mathbb{S}^1$ in $\mathbb{S}^3$.  In geometrical terms, the map $\pi: \mathbb{S}^3 \to \mathbb{S}^2$ makes $\mathbb{S}^3$ into the total space of a right principal fiber bundle with structure group $\mathbb{S}^1$ over $\mathbb{S}^2$.  We will refer to the orbits of the $\mathbb{S}^1$-action \eqref{circle_action} as the \emph{$\mathbb{S}^1$-fibers} of $\mathbb{S}^3$.

The Hopf map can be expressed conveniently in terms of the Pauli matrices as follows.
We let $\boldsymbol{\sigma}$ be the vector $(\sigma_1, \sigma_2, \sigma_3)$.  The Hopf map \eqref{projection} can then be described as 
\begin{equation} \label{hopf_pauli}
	\pi(\varphi) = \varphi^\dagger \boldsymbol{\sigma} \varphi.
\end{equation}
The right-hand side should be interpreted as a vector $\mathbf{x}$ in $\mathbb{R}^3$, whose components are given by $x^\alpha := \varphi^\dagger \sigma_\alpha  \varphi$, $\alpha = 1, 2, 3$.

The inner product of two vectors $\mathbf{x}, \mathbf{y} \in \mathbb{R}^3$ can be given a convenient description using the Hopf map.  Let $\mathbf{x} = \varphi^\dagger \boldsymbol{\sigma} \varphi$ and $\mathbf{y} = \psi^\dagger \boldsymbol{\sigma} \psi$.  A straightforward consequence of \eqref{pauli_comp} is then that
\begin{equation} \label{inner_product}
	\mathbf{x} \cdot \mathbf{y} = 2 (\varphi^\dagger \psi)(\psi^\dagger \varphi) - (\varphi^\dagger \varphi) (\psi^\dagger \psi).
\end{equation}

\paragraph{Connection one-form and curvature.}

On $\mathbb{S}^3$, there is a distinguished one-form $\mathcal{A}$ which will play a crucial role in obtaining the Lagrangian formulation for point vortices.  Explicitly, it is given by 
\[
	\mathcal{A}(\varphi) = \Im( \varphi^\dagger \, \mathbf{d} \varphi),
\]	
and we denote the contraction of $\mathcal{A}(\varphi)$ with a vector $\dot{\varphi} = (\dot{z}, \dot{u})$ by 
\begin{equation} \label{connection_one_form}
	\mathcal{A}(\varphi) \cdot \dot{\varphi} = 
	\Im( \varphi^\dagger \dot{\varphi}) 
	= \Im( z^\ast \dot{z} + u^\ast \dot{u}).
\end{equation}
Note the similarity between $\mathcal{A}$ and the one-form $\theta$ given in \eqref{complex_one_form}.

The form $\mathcal{A}$ is the connection one-form of a principal fiber bundle connection, but we will just treat it as a one-form.  The curvature of $\mathcal{A}$ is given by 
\[
	\mathbf{d} \mathcal{A} = 
	 \Im( \mathbf{d} \varphi^\dagger \wedge \mathbf{d} \varphi) =
	\Im (d{z}^\ast \wedge dz + d{u}^\ast \wedge du),
\]
and it can be shown that the area form $\Omega$ on $\mathbb{S}^2$, given by \eqref{area_form}, satisfies
\begin{equation} \label{pullback_quantization}
	\pi^\ast \Omega = 2 \mathbf{d} \mathcal{A}.
\end{equation}
This result states that the two-form $\Omega$, which is not exact, nevertheless becomes exact when pulled back along the Hopf map to $\mathbb{S}^3$.  This will allow us to construct a linear Lagrangian for point vortices on $\mathbb{S}^3$.

\section{Hamiltonian formulation of the vortex equations} \label{sec:hamiltonian}

In this section, we review the Hamiltonian description of the equations of motion \eqref{vortexequations} for point vortices on the unit sphere.   This system of first-order ODEs can be written in Hamiltonian form, where the phase space is the Cartesian product $(\mathbb{S}^2)^N$ of $N$ copies of the unit sphere $\mathbb{S}^2$, equipped with the symplectic form
\begin{equation} \label{sympform}
	\mathcal{B}_\Gamma(\mathbf{x}_1, \ldots, \mathbf{x}_N) = \sum_{i = 1}^N \Gamma_i \Omega(\mathbf{x}_i),
\end{equation}
where $\Omega$ is the standard symplectic area form on $\mathbb{S}^2$, given by \eqref{area_form}.

The Hamiltonian function is given by
\begin{equation} \label{hamiltonian_function}
	H = -\frac{1}{4\pi} \sum_{i < j} \Gamma_i \Gamma_j \log (2\sigma^2 + l_{ij}^2),
\end{equation}
where $l_{ij} := \left\Vert \mathbf{x}_i - \mathbf{x}_j \right\Vert$ is the chord distance between the $i$th and the $j$th vortex and $\sigma$ is the cutoff parameter introduced in \eqref{vortexequations}.

Hamilton's equations are then given by $\mathbf{i}_{\dot{\mathbf{x}}} \mathcal{B}_\Gamma = \mathbf{d} H$.  Explicitly, we are looking for a curve $t \mapsto \mathbf{x}(t) \in (\mathbb{S}^2)^N$ such that, for any variation $\delta \mathbf{x}(t) \in T_{\mathbf{x}(t)} (\mathbb{S}^2)^N$, we have that 
\[
	\mathcal{B}_\Gamma(\dot{\mathbf{x}}, \delta \mathbf{x}) = \mathbf{d} H(\mathbf{x}) \cdot \delta \mathbf{x}.
\]
Using the expression \eqref{area_form} for the symplectic form, this can be written as  
\[
	\sum_{i = 1}^N \Gamma_i \mathbf{x}_i ( \dot{\mathbf{x}}_i \times \delta \mathbf{x}_i) = \sum_{i = 1}^N \nabla_{\mathbf{x}_i}  H(\mathbf{x}) \cdot \delta \mathbf{x}_i, 
\]
so that 
\begin{equation} \label{ham_eq_multipliers}
	\Gamma_i  (\mathbf{x}_i \times \dot{\mathbf{x}}_i) = \nabla_{\mathbf{x}_i}  H(\mathbf{x}) + \lambda_i \mathbf{x}_i,
\end{equation}
where the Lagrange multipliers $\lambda_i$, $i = 1, \ldots, N$, have been introduced to enforce the constraint that the variations $\delta \mathbf{x}_i$ be tangent to the unit sphere, so that $\mathbf{x}_i \cdot \delta \mathbf{x}_i = 0$ for all $i = 1, \ldots, N$.  Taking the cross product of \eqref{ham_eq_multipliers} with $\mathbf{x}_i$ then results in 
\[
\Gamma_i  (\mathbf{x}_i \times \dot{\mathbf{x}}_i) \times \mathbf{x}_i = \nabla_{\mathbf{x}_i}  H(\mathbf{x}) \times \mathbf{x}_i, 
\]
and after expanding the double cross product and using the fact that $\left\Vert \mathbf{x}_i \right\Vert = 1$, we obtain 
\begin{equation} \label{hamiltonian_vector_equations}
	\Gamma_i \dot{\mathbf{x}}_i = \nabla_{\mathbf{x}_i}  H(\mathbf{x}) \times \mathbf{x}_i,
\end{equation}	
which is equivalent to \eqref{vortexequations}.

\section{Lagrangian formulation of the vortex equations on $\mathbb{S}^3$}

In this section, we show how the Hamiltonian equations \eqref{vortexequations} for point vortices can be given a Lagrangian formulation.  To do this, we lift the point vortex system from the two-sphere $\mathbb{S}^2$ to the three-sphere $\mathbb{S}^3$ using the Hopf fibration.

\paragraph{Pullback of the Hamiltonian $H$.}

Using the projection $\pi$ given in \eqref{projection}, we may pull back the Hamiltonian function on $\mathbb{S}^2$  to $\mathbb{S}^3$.  If we denote the Hamiltonian function \eqref{hamiltonian_function} by $H_{\mathbb{S}^2}$ and the pullback by $H_{\mathbb{S}^3}$, then we have that $H_{\mathbb{S}^3} = \pi^\ast H_{\mathbb{S}^2}$, or explicitly, 
\begin{equation} \label{pullback_hamiltonian}
	H_{\mathbb{S}^3}(\varphi_1, \ldots, \varphi_N) = H_{\mathbb{S}^2}( \pi(\varphi_1), \ldots, \pi(\varphi_N)),
\end{equation}
for all $\varphi_1, \ldots, \varphi_N \in \mathbb{S}^3$. Here, as in the remainder of the text, we have suppressed the dependence of $H_{\mathbb{S}^3}$ on the conjugate variables $\varphi_1^\dagger, \ldots, \varphi_N^\dagger$.

A straightforward computation shows that $H_{\mathbb{S}^3}$ is given by 
\begin{equation} \label{three_sphere_hamiltonian}
H_{\mathbb{S}^3}(\varphi_1, \ldots, \varphi_N) := -\frac{1}{4\pi} \sum_{i < j} \Gamma_i \Gamma_j \log\left[ 
		2\sigma^2 + 4(1 - \left| \left< \varphi_i, \varphi_j \right> \right|^2 ) \right].
\end{equation}
In the remainder, we will drop the subscript `$\mathbb{S}^3$' on the Hamiltonian function, denoting $H_{\mathbb{S}^3}$ simply as $H$.     Note that $H$ is invariant under multiplication by $e^{\mathrm{i} \theta} \in \mathbb{S}^1$ in each argument separately:
\begin{equation} \label{invarianceH}
	H(\ldots, e^{\mathrm{i}\theta} \varphi_k, \ldots)
	= H(\ldots,\varphi_k, \ldots),
\end{equation}
for $k = 1, \ldots, N$.  The infinitesimal version of this symmetry is 
\begin{equation} \label{infinitesimal_invarianceH}
\frac{\partial H}{\partial \varphi_k}\varphi_k - \varphi_k^\dagger \frac{\partial H}{\partial \varphi_k^\dagger} = 0,
\end{equation}
where there is \emph{no sum over the index $k$}.  

Since multiplying $\varphi_k$ by a phase factor $e^{\mathrm{i} \theta}$ corresponds to moving along the $\mathbb{S}^1$-fiber through $\varphi_k$, we have that $H$ depends only on the chord distance between the $\mathbb{S}^1$-fibers through $\varphi_1, \ldots, \varphi_N$.  This can be shown explicitly in \eqref{three_sphere_hamiltonian} by expressing the inner product $\left| \left< \varphi_i, \varphi_j \right> \right|$ in terms of the Euclidian distance $\mathcal{D}(\varphi_i, \varphi_j)$ between the $\mathbb{S}^1$-fibers through $\varphi_i$ and $\varphi_j$, where 
\[
	\mathcal{D}(\varphi_i, \varphi_j) = 2( 1 - \left| \left< \varphi_i, \varphi_j \right> \right|).
\]

\paragraph{The linear Lagrangian and the equations of motion.}  

We now have all the elements to formulate a Lagrangian description for point vortices using $\mathbb{S}^3$. Recall that a linear Lagrangian has the general form $L=\Theta-H$, where $\mathbf{d}\Theta$ is the symplectic form. The symplectic structure on $(\mathbb{S}^3)^N$ is given by the pullback of the symplectic structure on $(\mathbb{S}^2)^N$,
\begin{align*}
\pi^* \left(\sum_{i=1}^N \Gamma_i \Omega_i\right)&=\mathbf{d}\left(2\sum_{i=1}^N \Gamma_i \mathcal{A}_i\right),
\end{align*}
so it follows that $\Theta=2\sum_{i=1}^N \Gamma_i \mathcal{A}_i$. Therefore, we obtain
\begin{align} \label{linear_lagrangian}
	L   = 2 \sum_{i = 1}^N \Gamma_i \mathcal{A}(\varphi_i) \cdot \dot{\varphi}_i 
		- H(\varphi_1, \ldots, \varphi_N), 
\end{align}
where $\varphi_i \in \mathbb{S}^3$ for $i = 1, \ldots, N$. This generalizes the expression \eqref{plane_lagrangian} for the linear Lagrangian for point vortices in the plane.

The action functional is defined as 
\begin{equation} \label{action}
	\mathcal{S}(\varphi(\cdot)) = \int_{t_0}^{t_1} L(\varphi(t), \dot{\varphi}(t)) dt, 
\end{equation}
where $\varphi(t) := (\varphi_1(t), \ldots, \varphi_N(t))$ is a curve in $(\mathbb{S}^3)^N$ defined on the interval $[t_0, t_1]$, and its variation is given explicitly by 
\begin{equation} \label{continuous_vp}
	\delta \mathcal{S} = \sum_{i = 1}^N \delta \varphi_i^\dagger \left( - 2\mathrm{i} \Gamma_i\dot{\varphi}_i + \frac{\partial H}{\partial \varphi_i^\dagger} \right)
		+ \sum_{i = 1}^N \left( 2\mathrm{i} \Gamma_i\dot{\varphi}_i^\dagger + \frac{\partial H}{\partial \varphi_i}  \right) \delta \varphi_i,
\end{equation}
where the infinitesimal variations $\delta \varphi_i$ and $\delta \varphi_i^\dagger$ need to be prescribed carefully.  Since $\varphi_i$ is an element of $\mathbb{S}^3$, the variations $\delta \varphi_i$ are elements of $T_{\varphi} \mathbb{S}^3$.  Specifically, we have that $\delta \varphi_i$ is orthogonal to $\varphi_i$.  This relation may be incorporated using Lagrange multipliers $\lambda_i$, resulting in the Euler--Lagrange equations 
\begin{equation} \label{nonlinS}
	 \boxed{2\mathrm{i} \Gamma_i \dot{\varphi}_i =  \frac{\partial H}{\partial \varphi_i^\dagger} + \lambda_i \varphi_i,}
\end{equation} 
together with their Hermitian conjugates and the unit-length constraints 
\begin{equation} \label{unitlength}
	\left<\varphi_i,  \varphi_i \right> = 1.
\end{equation}
This equation is the analogue of \eqref{plane_equations} for vortices on $\mathbb{S}^3$ and can be seen as a \emph{nonlinear Schr\"odinger equation} on the product space $(\mathbb{S}^3)^N$.  The analogy with \eqref{plane_equations} can be made more striking by interpreting $\varphi_{i}$ as a \emph{unit quaternion}, so that \eqref{nonlinS} becomes (up to a constant) the quaternionic version of the complex equation \eqref{plane_equations}.

We will refer to the equations \eqref{nonlinS}, or one of their equivalent forms below, as the \emph{Hopf-lifted system on $(\mathbb{S}^3)^N$.}

\paragraph{Determining the Lagrange multipliers.}

A curious feature of these equations is that the multipliers $\lambda_i$ reflect \emph{gauge degrees of freedom}, that is, any choice of $\lambda_i$ will preserve the unit length constraint equally well.  To see this, take the time derivative of \eqref{unitlength} and substitute the equations of motion; the result is 
\[
	\frac{1}{2\mathrm{i}\Gamma_i} \left( - \frac{\partial H}{\partial \varphi_i} - \lambda_i \varphi_i^\dagger \right) \varphi_i + 
	\frac{1}{2\mathrm{i}\Gamma_i}  \varphi_i^\dagger \left(  \frac{\partial H}{\partial \varphi_i^\dagger} + \lambda_i \varphi_i \right) = 0,
\]
which simplifies to 
\[
	\frac{\partial H}{\partial \varphi_i}\varphi_i - \varphi_i^\dagger \frac{\partial H}{\partial \varphi_i^\dagger} = 0,
\]
from which $\lambda_i$ is absent.  This expression is nothing but the infinitesimal symmetry relation \eqref{infinitesimal_invarianceH} and is therefore identically satisfied.

With hindsight, it is not surprising that there is some indeterminacy in the solutions of \eqref{nonlinS}.  After all, these equations arise as pullbacks of equations on $\mathbb{S}^2$.  From this point of view, changing the multipliers $\lambda_i$ will change the dynamics in the direction of the $\mathbb{S}^1$-fibers, but will leave the horizontal dynamics (which projects down to $\mathbb{S}^2$) unchanged. 

This is similar to the \emph{un-reduction approach} of \cite{BrElHoGa2011}, in which a complicated dynamical system on a manifold $X$ is mapped into a simpler problem on the total space of a principal fiber bundle over $X$. Another conceptually related approach is presented in \cite{LeLeMc2009}, which considers continuous and discrete Lagrangian systems on $S^2$ by viewing $S^2$ as a homogeneous space with a transitive $SO(3)$ action, and lifting the Lagrangian on $S^2$ to $SO(3)$. This leads to a Lagrangian system on $SO(3)$ with non-isolated solutions parameterized by the isotropy subgroup, but a unique extremizing curve on $SO(3)$ can be obtained by restricting to horizontal curves with respect to a principle bundle connection. However, the projection of the curve onto $S^2$ is independent of the choice of the connection.

\paragraph{Relation with the equations on $(\mathbb{S}^2)^N$.}  

By construction, the flow of the equations \eqref{nonlinS} on $(\mathbb{S}^3)^N$ will project down onto the flow of the point vortex equations \eqref{vortexequations} on $(\mathbb{S}^2)^N$.  It is instructive, however, to see this explicitly.

We start again from the variational principle \eqref{continuous_vp}, but now we do not introduce a Lagrange multiplier to incorporate the unit-length constraint.  For the sake of clarity, we suppress the explicit index $i$ in \eqref{continuous_vp} to arrive at 
\begin{equation} \label{vp2}
	\delta \mathcal{S} =  \delta \varphi^\dagger \left( - 2\mathrm{i} \Gamma\dot{\varphi} + \frac{\partial H}{\partial \varphi^\dagger} \right)
		+ \left( 2\mathrm{i} \Gamma\dot{\varphi}^\dagger + \frac{\partial H}{\partial \varphi}  \right) \delta \varphi.
\end{equation}
As the variation $\delta \varphi$ is tangent to $\mathbb{S}^3$, it can be written as $\delta \varphi = A \varphi$,  where $A \in \mathfrak{su}(2)$; see \eqref{su_variations}. Similarly, we have that $\delta \varphi^\dagger = \varphi^\dagger A^\dagger = - \varphi^\dagger A$.  Upon substituting these expressions in \eqref{vp2}, we arrive at 
\begin{align*}
	\delta \mathcal{S} & = - \varphi^\dagger A \left( - 2\mathrm{i} \Gamma\dot{\varphi} + \frac{\partial H}{\partial \varphi^\dagger} \right)
		+ \left( 2\mathrm{i} \Gamma\dot{\varphi}^\dagger + \frac{\partial H}{\partial \varphi}  \right) A \varphi \\ 
		& = 2 \Re \left[ \left( 2\mathrm{i} \Gamma\dot{\varphi}^\dagger + \frac{\partial H}{\partial \varphi}  \right) A \varphi \right],
\end{align*} 
so that $\delta \mathcal{S} = 0$ for all $A \in \mathfrak{su}(2)$ if and only if 
\begin{equation} \label{projected_eom}
	\boxed{\Re \left[ \left( 2\mathrm{i} \Gamma\dot{\varphi}^\dagger + \frac{\partial H}{\partial \varphi}  \right) \mathrm{i} \sigma_\alpha \varphi \right] = 0, \quad 
		\alpha = 1, 2, 3,}
\end{equation}
where the $\sigma_\alpha$ are the Pauli matrices \eqref{pauli}.  Note that these equations are equivalent to \eqref{nonlinS}.

We now let $\mathbf{x} \in \mathbb{S}^2$ be the image of $\varphi \in \mathbb{S}^3$ under the Hopf map, and we recall from \eqref{hopf_pauli} that the components of $\mathbf{x}$ are given by $x_\alpha = \varphi^\dagger \sigma_\alpha \varphi$.  Taking the time derivative, we obtain 
\begin{equation} \label{derivative}
	\dot{x}_\alpha = 2 \Re \left( \dot{\varphi}^\dagger  \sigma_\alpha \varphi \right).
\end{equation}
Similarly, we recall that the Hamiltonian functions $H_{\mathbb{S}^2}$ and $H_{\mathbb{S}^3}$ are related by \eqref{pullback_hamiltonian}, or explicitly $H_{\mathbb{S}^3}(\varphi) = H_{\mathbb{S}^2}(\varphi^\dagger \sigma_\alpha \varphi)$.
Taking the derivative with respect to $\varphi$ yields
\begin{equation} \label{der_relation}
	\frac{\partial H_{\mathbb{S}^3}}{\partial \varphi} = \frac{\partial H_{\mathbb{S}^2}}{\partial x_\beta} \varphi^\dagger \sigma_\beta, 
\end{equation}
and a small calculation, involving the multiplication identity \eqref{pauli_mult}, then shows that 
\[
	\Re \left[ \mathrm{i} \frac{\partial H}{\partial \varphi} \sigma_\alpha \varphi \right] = 
		\sum_{\beta, \gamma} \epsilon_{\alpha \beta \gamma} \frac{\partial H_{\mathbb{S}^2}}{\partial x_\beta} x_\gamma 
		= (\nabla_{\mathbf{x}} H_{\mathbb{S}^2} \times \mathbf{x})_\alpha.
\]
Substituting this expression and \eqref{derivative} into \eqref{projected_eom} then results in the following vector equations: $\Gamma \dot{\mathbf{x}} =  \nabla_{\mathbf{x}} H_{\mathbb{S}^2} \times  \mathbf{x}$, which, upon restoring the sum over all vortices, are nothing but the point vortex equations \eqref{hamiltonian_vector_equations}.

\paragraph{Singular vorticity distributions on $\mathbb{S}^3$.}

The solutions of the equations \eqref{nonlinS} on $(\mathbb{S}^3)^N$ project down onto the solutions of the point vortex equations on $(\mathbb{S}^2)^N$. One can therefore view the equations \eqref{nonlinS} on $(\mathbb{S}^3)^N$ as describing singular vorticity fields supported along the $\mathbb{S}^1$-fibers of the Hopf fibration. These fibers are well known to be pairwise linked, but we do not know of any consequences of this fact for the dynamics of point vortices.

The interpretation of singular distributions of vorticity supported along fibers of the Hopf fibration agrees with the results of \cite{Sh2012, Kh2012}, who show that a singular vorticity distribution must necessarily be of codimension 2 or less.

\paragraph{Pre-symplectic formulation of the lifted equations.}

In this concluding paragraph, we show how the equations \eqref{nonlinS} on $(\mathbb{S}^3)^N$ can be written in pre-symplectic form, and we finish with some remarks on the relation between the indeterminacy of the Lagrange multipliers in \eqref{nonlinS} and the appearance of gauge freedom.  The presymplectic point of view is useful to shed further light on the nature of the equations \eqref{nonlinS}, but we will not use it in the remainder of the paper. This paragraph can therefore be omitted on a first reading.

We recall first of all that, given a Lagrangian $L : TQ \to \mathbb{R}$, the Euler-Lagrange equations can be written intrinsically as 
\begin{equation} \label{lagrangian_eom}
  \mathbf{i}_X \Omega_L = \mathbf{d} E_L,
\end{equation}
where $\Omega_L$ is the pullback $(\mathbb{F}L)^\ast \Omega$ of the canonical symplectic form $\Omega$ on $T^\ast Q$ along the Legendre transformation $\mathbb{F}L$, and $E_L$ is the Lagrangian energy, defined as $E_L(q, v) = \left< v, \mathbb{F}L(q, v) \right> - L(q, v)$. Below, instead of pulling everything back by the Legendre transform, we will work directly on the \emph{primary constraint submanifold}, defined as the image of the Legendre transform in $T^\ast Q$. 

For the Hopf system, we begin by calculating the Legendre transform $\mathbb{F}L : T(\mathbb{S}^3)^N \to T^\ast (\mathbb{S}^3)^N$. This map is given by
\[
  \mathbb{F}L : (\varphi_i, \dot{\varphi}_i) \mapsto ( \varphi_i, \pi_i), 
    \quad \text{where} \quad \pi_i = \frac{\partial L}{\partial \dot{\varphi}_i}
        = 2 \Gamma_i \mathcal{A}(\varphi_i).
\]
The primary constraint submanifold is the image of $\mathbb{F} L$ and is clearly seen to be a submanifold of $T^\ast (\mathbb{S}^3)^N$ which is diffeomorphic to $(\mathbb{S}^3)^N$. For the pullback of the canonical symplectic form on $ T^\ast (\mathbb{S}^3)^N$ to the primary constraint submanifold we now obtain
\[
  \mathcal{B}_{(\mathcal{S}^3)^N} 
    = \sum_{i=1}^N \mathbf{d} \pi_i \wedge \mathbf{d} \varphi_i 
    = 2 \sum_{i=1}^N \Gamma_i \, \mathbf{d} \mathcal{A}(\varphi_i) 
      \wedge \mathbf{d} \varphi_i.
\]
Note that, as the notation suggests, $\mathcal{B}_{(\mathbb{S}_3)^N}$ is the pullback to $(\mathbb{S}^3)^N$ of the point vortex symplectic form $\mathcal{B}$ given in \eqref{sympform}: $\mathcal{B}_{(\mathbb{S}_3)^N} = \pi^\ast \mathcal{B}$, with $\pi: \mathbb{S}^3 \to \mathbb{S}^2$ the Hopf map.  As a result, $\mathcal{B}_{(\mathbb{S}_3)^N}$ is a pre-symplectic form: its kernel consists of all vectors which are tangent to the fibers of the Hopf fibration, and a small calculation shows that 
\[
  \mathrm{ker} \, \mathcal{B}_{(\mathbb{S}_3)^N} = \mathrm{span} \, \left\{ 
        \frac{\partial}{\partial \varphi_k} \varphi_k - 
        \varphi_k^\dagger \frac{\partial}{\partial \varphi_k^\dagger}, \quad
        k = 1, \ldots, N
      \right\}.
\]

Furthermore, it can easily be checked that the Lagrangian energy $E_L$
induces a function on the primary constraint submanifold which is
nothing but the lifted Hamiltonian function $H_{\mathbb{S}^3}$. The Euler-Lagrange equations \eqref{lagrangian_eom} now become
\begin{equation} \label{ham_eq}
	\mathbf{i}_{\Gamma} \mathcal{B}_{(\mathbb{S}_3)^N} = \mathbf{d} H_{\mathbb{S}_3}. 
\end{equation}
These equations do not determine the dynamics completely: given any
solution $\Gamma$ of \eqref{ham_eq}, we may add to it an arbitrary
element of $\mathrm{ker} \, \mathcal{B}_{(\mathbb{S}_3)^N}$ without
changing the physical degrees of freedom. In the literature on
degenerate Lagrangians (see for instance \cite{Go1979}), the elements
of $\mathrm{ker} \, \mathcal{B}_{(\mathbb{S}_3)^N}$ are referred to as
\emph{gauge vector fields} for precisely this reason. As the gauge
vector fields generate in this case a flow along the fibers of the
Hopf fibration, we have that the physical degrees of freedom take
values in the quotient space $(\mathbb{S}^3/\mathbb{S}^1)^N \cong
(\mathbb{S}^2)^N$. This is of course nothing but a restatement of the
fact that the Hopf lifted system arose by lifting point vortex
dynamics from $\mathbb{S}^2$ to $\mathbb{S}^3$. 

Lastly, we may resolve the issue of gauge indeterminacy by replacing $\mathbb{S}^3$ by its \emph{symplectification}, which is the product manifold $\mathbb{S}^3 \times \mathbb{R}^+$ equipped with the symplectic form 
\[
	\tilde{\mathcal{B}} = \mathbf{d} ( r \mathcal{A}),
\]
where $r$ is the coordinate on the $\mathbb{R}^+$-factor.  The motion of point vortices on this twice-enlarged space projects down to the motion of point vortices on $\mathbb{S}^2$, and can be viewed, paraphrasing the terminology of \cite{Ko2005}, as a version of ``prequantum vortex dynamics.''\footnote{We thank M. Gotay for bringing this point to our attention.}

\section{Variational integrators on $SU(2)^N$}

In this section, we propose a discrete version of the Hopf-lifted system on $(\mathbb{S}^3)^N$.  We begin by discretizing the linear Lagrangian \eqref{linear_lagrangian} using centered finite differences.  By taking discrete variations, we then obtain a discrete version of the equations \eqref{nonlinS} where the constraints are enforced using a Lagrange multiplier.  These equations can be seen as a version of the Moser-Veselov equations (see \cite{MoVe1991}) on $(\mathbb{S}^3)^N$. 

By projecting onto the annihilator space of the constraint forces, we then obtain a discrete version of the projected equations \eqref{projected_eom}.  Finally, we use the isomorphism between $\mathbb{S}^3$ and $SU(2)$ to write the discrete equations of motion in the form of a homogeneous space variational integrator (see \cite{LeLeMc2009}) on $SU(2)$ and we argue that this form of the equations is especially well-suited for numerical implementation.

\subsection{Discrete Lagrangian and discrete equations of motion}

Let $M$ be the number of discrete time steps, with constant time increment $h > 0$, and denote the variables at time $t_n := n h$ by $\varphi^n := (\varphi^n_1, \ldots, \varphi^n_N) \in (\mathbb{S}^3)^N$.  We now propose a discrete counterpart of the linear Lagrangian $L$ in \eqref{linear_lagrangian} by approximating the action functional \eqref{action} over the interval $[t_n, t_{n+1}]$ by using piecewise linear interpolants and the midpoint rule (see \cite{MaWe2001}).  In this way, we construct a discrete Lagrangian $L_d : (\mathbb{S}^3)^N \times (\mathbb{S}^3)^N \to \mathbb{R}$ of the form
\begin{equation} \label{discrete_lagrangian}
	L_d(\varphi^n, \varphi^{n+1}) = h L \left(  \frac{\varphi^n + \varphi^{n+1}}{2}, \frac{\varphi^{n+1} - \varphi^n}{h} \right). 
\end{equation}
Explicitly,  the discrete linear Lagrangian is given by
\begin{align} \label{linear_lagrangian_discrete}
	L_d(\varphi^n, \varphi^{n+1}) = & \,  2 \sum_{i = 1}^N \Gamma_i \mathcal{A}(\varphi_i^{n+1/2}) \cdot (\varphi_i^{n+1} - \varphi_i^n) 
		- h H(\varphi^{n+1/2}),
\end{align}
where $\varphi^{n+1/2} := (\varphi^n + \varphi^{n+1})/2$.

The discrete action sum is now defined as 
\[
	\mathcal{S}_d(\varphi^1, \ldots, \varphi^M) = \sum_{i = 1}^{M-1} L_d(\varphi^n, \varphi^{n+1}),
\]
and taking variations with respect to $\varphi^n_i$ and $({\varphi}^n_i)^\dagger$ yields
\begin{align}
	\delta \mathcal{S}_d = \sum_{k = 1}^N \sum_{n=1}^{M-1}  \delta (\varphi^n_k)^\dagger \Big[ 	
	& -\mathrm{i} \Gamma_k \left( {\varphi}^{n+1}_k - {\varphi}^{n-1}_k \right) - h \lambda^n_k {\varphi}^n_k \nonumber \\
		& + \frac{h}{2}\left( D_{\varphi^\dagger_k}H(\varphi^{n-1/2}) + D_{\varphi^\dagger_k}H(\varphi^{n+1/2}) \right) 
		 \Big]   + \text{(c.c.)}, \label{discrete_vp}
\end{align}
where ``(c.c.)'' stands for the complex conjugate of the expression preceding it. Here, and in the remainder of the paper, $D_{\varphi_k}$ denotes the derivative with respect to $\varphi_k$, and similarly for $D_{\varphi_k^\dagger}$.

   The discrete Euler--Lagrange equations are hence 
\begin{equation} \label{discrete_EL}
\boxed{-\mathrm{i} \Gamma_k \left( {\varphi}^{n+1}_k - {\varphi}^{n-1}_k \right)  
		 + \frac{h}{2}\left( D_{\varphi^\dagger_k}H(\varphi^{n-1/2}) + D_{\varphi^\dagger_k}H(\varphi^{n+1/2}) \right) 
		 - h \lambda^n_k {\varphi}^n_k = 0,}
\end{equation}
together with their Hermitian conjugates and the unit-length constraints 
\begin{equation} \label{disc_unit_length}
	\left<\varphi_i^{n+1},  \varphi_i^{n+1} \right> = 1,
\end{equation}
and can be viewed as the discrete analogues of the continuous equations \eqref{nonlinS}.

In contrast to the continuous case, the Lagrange multipliers $\lambda^n_k$ in \eqref{discrete_EL} are no longer arbitrary. Instead, they can be found by substituting the discrete equations of motion into the unit-length constraint \eqref{disc_unit_length} and solving the resulting system of quadratic equations for $\lambda^n_k$.

\paragraph{Other discretizations.}  

Instead of the midpoint quadrature formula, leading to the discrete Lagrangian \eqref{discrete_lagrangian}, we could have chosen another approximation for the discrete Lagrangian, leading to a different set of discrete equations.  Although these equations formally exhibit the same properties (symplecticity, preservation of the unit length constraint, etc.) as the midpoint method introduced previously, some of them suffer from undesirable side-effects.  A particularly revealing example is obtained by using the trapezoid rule instead of the midpoint rule, leading to the discrete Lagrangian
\[
	L_d(\varphi^n, \varphi^{n+1}) = \frac{h}{2}\left( L \left(  \varphi^n, \frac{\varphi^{n+1} - \varphi^n}{h} \right) + 
		  L \left(  \varphi^{n+1}, \frac{\varphi^{n+1} - \varphi^n}{h} \right) \right), 
\]
which gives rise to the discrete Euler--Lagrange equations 
\begin{equation} \label{sphere_eq_unstable}
-\mathrm{i} \Gamma_k \left( {\varphi}^{n+1}_k - {\varphi}^{n-1}_k \right)  
		 + h D_{{\varphi}^\dagger_k}H(\varphi^{n})  
		 = h \lambda^n_k {\varphi}^n_k,
\end{equation}
together with unit-length constraints \eqref{disc_unit_length}.  This system, however, is equivalent to the 2-step symmetric explicit midpoint integrator (see \cite[Sec.~XV.3.2]{HaLuWa2002}), which is well-known to exhibit parasitic solutions which grow linearly in time.   The same observation is true for the integrator of \cite{RoMa2002} with $\sigma = 0$ or $1$;\footnote{Here $\sigma$ refers to the interpolation parameter used in \cite{RoMa2002}, and should not be confused with the cutoff parameter used in the rest of the current paper.}  see appendix~\ref{sec:planar} for details.

\subsection{The projected midpoint equations} \label{sec:projected_midpoint}

In this section, we show how the discrete Euler-Lagrange equations \eqref{discrete_EL} can be significantly simplified under a number of modest assumptions.  The process of simplifying the equations proceeds as follows:
\begin{enumerate}
	\item We first show how the discrete Euler-Lagrange equations can be written in projected form, i.e. without reference to the Lagrange multipliers. The result is \eqref{projected_discrete} below.
	\item We then show how this discrete second-order equation may be written as the composition of two first-order equations which are mutually adjoint, resulting in \eqref{projected_discrete2}, \eqref{LHS}.
	\item Lastly, in the case of a $\mathbb{S}^1$-invariant Hamiltonian, we show that both first-order equations can be solved by solving a simple implicit midpoint method on $\mathbb{S}^3$, as in \eqref{implicit_midpoint_method_S3}.
\end{enumerate}

The first two simplifications can be made for any Hamiltonian system on $\mathbb{S}^3$. The last simplification can be only be made when the Hamiltonian on $\mathbb{S}^3$ is $\mathbb{S}^1$-invariant (and hence is the pullback of a Hamiltonian function on $\mathbb{S}^2$). This is the case for point vortex dynamics and in fact for the majority of physical systems on $\mathbb{S}^3$ (such as spin-1/2 systems), and is therefore not a very restrictive assumption.

\paragraph{First simplification: the projected Euler-Lagrange equations.}
To obtain an equivalent version of the equations \eqref{discrete_EL} which does not involve the Lagrange multipliers $\lambda^n_k$, we return to the discrete variational principle \eqref{discrete_vp}, given by 
\[
	\delta \mathcal{S}_d =  \sum_{n=1}^{M-1}  \delta ({\varphi}^n)^\dagger \Big[ 	
	 -\mathrm{i} \Gamma \left( {\varphi}^{n+1} - {\varphi}^{n-1} \right) 
		 + \frac{h}{2}\left( D_{{\varphi}^\dagger}H(\varphi^{n-1/2}) + D_{{\varphi}^\dagger}H(\varphi^{n+1/2}) \right) 
		 \Big]   + \text{(c.c.)},
\]
where we have suppressed the vortex index $k$.  Instead of introducing a Lagrange multiplier to enforce the unit-length constraint, we impose the condition that the variations $\delta \varphi$ and $\delta \varphi^\dagger$ are tangent to the sphere, so that they can be written as 
\[
	\delta \varphi = A \varphi, \quad \text{and} \quad \delta \varphi^\dagger = -\varphi^\dagger A, 
\]
where $A \in \mathfrak{su}(2)$ is arbitrary; see \eqref{su_variations}.  Similar constrained variations were adopted in \cite{LeLeMc2009}. Proceeding as in the case of the continuous variational principle, we then arrive at the following discrete equations: 
\begin{equation} \label{projected_discrete}
	\boxed{\Re \left[ (\varphi^n)^\dagger (\mathrm{i} \sigma_\alpha) \left(-\mathrm{i} \Gamma \left( {\varphi}^{n+1} - {\varphi}^{n-1} \right) 
		 + \frac{h}{2}\left( D_{{\varphi}^\dagger}H(\varphi^{n-1/2}) + D_{{\varphi}^\dagger}H(\varphi^{n+1/2}) \right) \right)  \right] = 0,}
\end{equation}
for $\alpha = 1, 2, 3$.  These equations are supplemented by the unit-length constraint \eqref{disc_unit_length}.  Note that the equations \eqref{projected_discrete} are the discrete version of the continuous projected vortex equations \eqref{projected_eom}.  Another way to arrive at these equations is simply to project the discrete Euler--Lagrange equations \eqref{discrete_EL} onto the subspace orthogonal to $\varphi^n$, which is equivalent to applying the discrete null space method of \cite{LeMaOr2008}.

\paragraph{Second simplification: the equivalent first-order system.}   The equations \eqref{projected_discrete} are second-order discrete equations: given $(\varphi^{n-1}, \varphi^n)$, the equations can be solved to find $\varphi^{n+1}$.   The discrete equations of motion are hence not self-starting: given the initial positions $\varphi^0$ for the point vortices, a standard one-step integrator needs to be used to find the positions $\varphi^1$ at the intermediate time $t_1 = h$.  Afterwards, the discrete equations of motion can be used to integrate the system forwards in time.

In the next paragraph, we will find a way to recast the second-order equations as an equivalent first-order system, which is self-starting. We begin by writing the two-step method \eqref{projected_discrete} as the composition of two one-step methods, which turn out to be mutually adjoint.  That this decomposition is possible is a consequence of the fact that the Lagrangian \eqref{linear_lagrangian} is linear in the velocities, and will be analyzed further in Appendix~\ref{sec:planar}.  For now, we just focus on the computations for the point vortex equations.

We first write the equations \eqref{projected_discrete} as 
\begin{equation} \label{projected_discrete2}
	\Re \left[ (\varphi^n)^\dagger (\mathrm{i} \sigma_\alpha) \left(-\mathrm{i} \Gamma ( {\varphi}^{n+1} - {\varphi}^{n} ) 
		 + \frac{h}{2} D_{{\varphi}^\dagger}H(\varphi^{n+1/2})\right)  \right] =  - d^n_\alpha,
\end{equation}
where the \emph{slack variables} $d^n_\alpha$ depend on the configurations $\varphi^n$ and $\varphi^{n-1}$ only and are given by
\begin{equation} \label{LHS}
	d^n_\alpha := \Re \left[ (\varphi^n)^\dagger (\mathrm{i} \sigma_\alpha) \left(-\mathrm{i} \Gamma ( {\varphi}^{n} - {\varphi}^{n-1} ) 
		 + \frac{h}{2} D_{{\varphi}^\dagger}H(\varphi^{n-1/2})\right)  \right]. 
\end{equation}

One way of solving these equations is to start with initial conditions $(\varphi^{n-1}, \varphi^n)$, compute the slack variables $d^n_\alpha$ from \eqref{LHS}, and to find $\varphi^{n + 1}$ from \eqref{projected_discrete2}. We will discuss this approach further in Appendix~\ref{app:no_invariance}, but below we discuss an important simplification which can be made whenever the Hamiltonian $H$ is $\mathbb{S}^1$-symmetric.

\paragraph{Final simplification: the equivalence with the implicit midpoint method.}

For the point vortex system, one important further simplification can be made. It turns out that the system \eqref{projected_discrete2}, \eqref{LHS} of first-order equations, with $d^n_\alpha = 0$, is equivalent to the \emph{implicit midpoint method} $\Psi_h: \varphi^n \mapsto \varphi^{n+1}$, where $\varphi^{n+1}$ is given by 
\begin{equation} \label{implicit_midpoint_method_S3}
	\boxed{-\mathrm{i} \Gamma ( {\varphi}^{n+1} - {\varphi}^{n} ) 
		 + \frac{h}{2} D_{{\varphi}^\dagger}H(\varphi^{n+1/2}) = 0.}
\end{equation}
As this method is extremely easy to implement, this is a drastic improvement over the previous formulations of the discrete Euler-Lagrange equations. As we shall see below, this approach is only applicable if the Hamiltonian $H$ is $\mathbb{S}^1$-invariant, as is the case for point vortices. In Appendix~\ref{app:no_invariance}, we sketch an alternative approach to deal with the case of Hamiltonians that are not invariant.

A first noteworthy property of the midpoint method \eqref{implicit_midpoint_method_S3} is that it is length-preserving, without the need for spurious projections. To see this, multiply both sides of the equation by $\mathrm{i} (\varphi^{n+1/2})^\dagger$, and take the real part to obtain 
\[
	\Gamma \Re \left[ (\varphi^{n+1/2})^\dagger ({\varphi}^{n+1} - {\varphi}^{n}) \right]
		=  - \frac{h}{2} \Re \left[ (\varphi^{n+1/2})^\dagger 
			D_{{\varphi}^\dagger}H(\varphi^{n+1/2}) \right].
\]
The right-hand side of this equation vanishes because of the $\mathbb{S}^1$-symmetry invariance property \eqref{infinitesimal_invarianceH}, and we are left with 
\begin{align*}
	0 = \Gamma \Re \left[ (\varphi^{n+1/2})^\dagger 
		({\varphi}^{n+1} - {\varphi}^{n}) \right] & = 
		\frac{\Gamma}{2} \left( 
			  \left\Vert \varphi^{n+1} \right\Vert^2 	
			- \left\Vert \varphi^n \right\Vert^2
		\right),
\end{align*}
which shows that the method is length-preserving.  We summarize this in the following proposition.

\begin{proposition}
If the Hamiltonian $H$ is $\mathbb{S}^1$-invariant in each of its arguments (i.e.  \eqref{infinitesimal_invarianceH} holds), then the implicit midpoint method \eqref{implicit_midpoint_method_S3} is length-preserving.
\end{proposition}

To prove the equivalence between the implicit midpoint method \eqref{implicit_midpoint_method_S3} and the first-order equations \eqref{projected_discrete2}, \eqref{LHS} with $d^n_\alpha = 0$, we proceed as follows. It is clear that a solution of \eqref{implicit_midpoint_method_S3} is a solution of \eqref{projected_discrete2} and \eqref{LHS} with $d^n_\alpha = 0$, since the latter are just the projection of the implicit midpoint method on the tangent spaces at $\varphi^{n}$ and $\varphi^{n+1}$, respectively.

To prove the converse, we assume first that $(\varphi^{n-1}, \varphi^n) \in \mathbb{S}^3 \times \mathbb{S}^3$ is a solution of the equation \eqref{LHS} with $d^n_\alpha = 0$. This is equivalent to 
\[
-\mathrm{i} \Gamma ( {\varphi}^{n} - {\varphi}^{n-1} ) 
		 + \frac{h}{2} D_{{\varphi}^\dagger}H(\varphi^{n-1/2}) = \lambda \varphi^n,
\]
for some real-valued Lagrange multipliers $\lambda$. We now again multiply both sides by $\mathrm{i} (\varphi^{n+1/2})^\dagger$ and take the real part. After performing essentially the same manipulations as before, we then arrive at 
\[
	C \lambda = \frac{\Gamma}{2} ( \left\Vert \varphi^{n} \right\Vert^2 -
		\left\Vert \varphi^{n-1} \right\Vert^2 ) = 0,
\]
where $C = \Re \left[\mathrm{i} (\varphi^{n+1/2})^\dagger \varphi^1 \right]$. As $C \ne 0$, this implies that $\lambda = 0$ so that $(\varphi^{n-1}, \varphi^n)$ solves the implicit midpoint method \eqref{implicit_midpoint_method_S3}. The same approach can also be used to show that the solutions of \eqref{projected_discrete2} coincide with the solutions of the implicit midpoint method \eqref{implicit_midpoint_method_S3}.

\begin{proposition}
If the Hamiltonian $H$ is $\mathbb{S}^1$-invariant in each of its arguments, then the solutions of the implicit midpoint method \eqref{implicit_midpoint_method_S3} coincide with the solutions of the projected equations \eqref{projected_discrete2}, \eqref{LHS}.
\end{proposition}

\paragraph{Aside: adjointness of the first-order equations.} 

The first-order equations \eqref{projected_discrete2} and \eqref{LHS} share a particular structure with other methods derived from linear Lagrangians, as we shall see in Appendix~\ref{sec:planar}. We now discuss some of this structure but as the remainder of this paragraph does not affect the development of the variational integrator, it can safely be omitted on a first reading.

By viewing \eqref{projected_discrete2} as an equation for $\varphi^{n+1}$ we may introduce a map $\Phi_h : \mathbb{S}^3 \to \mathbb{S}^3$ defined by the property that $\Phi_h(\varphi^n) = \varphi^{n+1}$ if and only $(\varphi^n, \varphi^{n+1})$ satisfies \eqref{projected_discrete2}, where the slack variables $d^n_\alpha$ are viewed as parameters.   Likewise, \eqref{LHS} can be viewed as an equation for $\varphi^{n}$ given $\varphi^{n-1}$, and we let $\Psi_h : \mathbb{S}^3 \to \mathbb{S}^3$ be the map which takes $\varphi^{n-1}$ into $\varphi^n$. 

A small calculation then shows that $\Phi_h$ and $\Psi_h$ are each other's adjoint, that is, 
\[
	\Psi_h(\varphi^{n-1}) = \varphi^n  \quad \text{if and only if} \quad \Phi_{-h}(\varphi^n) = \varphi^{n-1}.
\]
The full discrete Euler-Lagrange equations \eqref{projected_discrete} can therefore be solved by composing the one-step methods $\Psi_h$ and $\Psi_h^\ast = \Phi_h$. As these methods are each others adjoint, the result is symmetric and guaranteed to be of second order. In Appendix~\ref{sec:planar} we argue that this decomposition of a two-step method into a system of adjoint one-step methods is a general feature of numerical methods derived from a linear Lagrangian.

\subsection{Properties of the variational integrator}
   
\paragraph{The symplectic form.}
Since we have started from the midpoint discretization of a continuous Hamiltonian, the resulting integrator will be second-order accurate, symplectic, and (by construction) unit-length preserving (see \cite{MaWe2001}).  The symplectic form preserved by the numerical algorithm is not the weighted area form on $(\mathbb{S}^2)^N$ given in \eqref{sympform} but the two-form 
\[
	\Im \left(\frac{\partial^2 L_d}{\partial (\varphi_k^{(0)})^\dagger 
		\partial \varphi_k^{(1)}}  \, 
		d  (\varphi_k^{(0)})^\dagger \wedge d \varphi_k^{(1)}
		\right),
\]
where $L_d$ is the discrete Lagrangian given in \eqref{linear_lagrangian_discrete}, which is an $\mathcal{O}(h)$ perturbation of it (see also \cite{RoMa2002}).  The fact that the integrator is symplectic explains --- through backward error analysis (see, for example, \cite{BeGi1994, Ha1994, HaLu1997, Re1999}) --- its good near-energy preservation properties.

\paragraph{The moment of vorticity.} The discrete Lagrangian \eqref{linear_lagrangian_discrete} has a symmetry which has gone unnoticed up to this point: if we translate $\varphi^0$ and $\varphi^1$ by the same element $U \in SU(2)$, then the Lagrangian stays invariant:
\[
	L_d(U \varphi^0, U \varphi^1) = L_d(\varphi^0, \varphi^1),
		\quad \text{for all $U \in SU(2)$}.
\]
This symmetry gives rise via Noether's theorem to a conserved quantity $\mathbf{J}$, whose components are given by 
\begin{align*}
	J_\alpha(\varphi^0, \varphi^1) & 
		= \Re \left( D_1 L(\varphi^0, \varphi^1)^\dagger 
			(\mathrm{i} \sigma_\alpha \varphi^0) \right)\\
		& = \Re \left( \left( \mathrm{i} \Gamma (\varphi^1)^\dagger + 
		\frac{h}{2} D_{\varphi} H(\varphi^{1/2}) \right) 
			\mathrm{i} \sigma_\alpha \varphi^0  \right),
\end{align*}
where the first expression is the standard expression for the discrete momentum map, see e.g. \cite{MaWe2001}, and $D_1 L$ refers to the derivative of the Lagrangian with respect to the first argument.

The conserved quantity $\mathbf{J}$ can be rewritten further by taking the complex conjugate of the term inside the brackets and writing 
\[
	J_\alpha = - \Re \left( (\varphi^0)^\dagger \mathrm{i} \sigma_\alpha 
		\left( -\mathrm{i} \Gamma ( \varphi^1- \varphi^0) + 
		\frac{h}{2} D_{\varphi^\dagger} H(\varphi^{1/2}) \right) 
			  \right) + 
	\Gamma \Re \left( (\varphi^0)^\dagger \sigma_\alpha \varphi_0 \right).
\]
An important simplification now occurs: the first term involves the expression \eqref{implicit_midpoint_method_S3} for the discrete equations of motion to $J_\alpha$, and hence vanishes along the trajectories of the equations of motion. We are left with 
\[
	J_\alpha = 
	\Gamma \Re \left( (\varphi^0)^\dagger \sigma_\alpha \varphi_0 \right) 
	= \Gamma x^0_\alpha, 
\]
where $\mathbf{x}_0 \in \mathbb{R}^3$ is the projection under the Hopf fibration of $\varphi_0$.

We summarize these developments in the following proposition, where we have restored the index $k$ labeling the individual vortices.

\begin{proposition}
Along the solutions of the discrete equations of motion \eqref{implicit_midpoint_method_S3}, the \emph{moment of vorticity} 
\[
	\mathbf{J} = \sum_{k = 1}^N \Gamma_k \mathbf{x}_k 
\]
is exactly preserved.
\end{proposition}

\section{Numerical results}

Throughout this section, we will compare the behavior of the Hopf integrator with a number of other integrators:
\begin{enumerate}
	\item A standard, non-variational integrator: we chose  a standard explicit 4th-order Runge--Kutta method (RK4), composed with projection onto the unit sphere in order to preserve the unit-length constraint.  It is well known (see e.g. \cite{HaLuWa2002}) that the resulting method is still 4th-order in time. For the numerical order comparisons in Section~\ref{sec:polvani_dritschel} we instead use the projected version of Heun's method, which we labeled as RK2.

	\item The implicit midpoint method on $\mathbb{S}^2$, given by 
	\begin{equation} \label{midpoint_algo}
		\frac{\mathbf{x}^{n+1}_k - \mathbf{x}^{n}_k}{h} =  
		\frac{1}{4 \pi} \sum_{j \ne k} \Gamma_j  \frac{ \mathbf{x}^{n+1/2}_j \times  \mathbf{x}^{n+1/2}_k}{1 + \sigma^2 -  \mathbf{x}^{n+1/2}_k \cdot  \mathbf{x}^{n+1/2}_j}.
	\end{equation}
	This is just the standard midpoint method, applied to the  vortex equations \eqref{vortexequations}. Note that for this vector field, the implicit midpoint method in fact stays on the unit sphere without the need for an explicit projection.  To see this, take the dot product of both sides of the equation with $\mathbf{x}^{n+1/2}_k$ and observe that the right-hand side vanishes, so that we get 
	\[
		(\mathbf{x}^{n+1}_k - \mathbf{x}^{n}_k) \cdot \mathbf{x}^{n+1/2}_k = 0, 
	\]
	which is equivalent to $\left\Vert \mathbf{x}^{n}_k \right\Vert = \left\Vert \mathbf{x}^{n+1}_k \right\Vert$, i.e., the length is preserved.  This is not a general feature of the midpoint method, but is a consequence of the particular form of the point vortex equations.  Furthermore, the midpoint method is symmetric under the interchange $\mathbf{x}^{n}_k \leftrightarrow \mathbf{x}^{n+1}_k$, $h \leftrightarrow -h$, and as a result the method seems to have good long-term conservation properties. 
	
We note that, despite the similarities, the midpoint method on the sphere is not exactly equal to the Hopf method. For the midpoint method, the gradient of the Hamiltonian is evaluated at the midpoint $\mathbf{x}^{n + 1/2}$, which is not exactly on the surface of the sphere, whereas for the Hopf integrator, the gradient is evaluated at the projection $\pi( \varphi^{n + 1/2})$ which is on the surface of the sphere.
	
	\item The Lie--Poisson method of \cite{EnFa2002}, applied to the point vortex equations of Section~\ref{sec:hamiltonian}.  This second-order method preserves the vortex moment explicitly, and is a self-adjoint Lie group method, resulting in bounded energy error.  However, this method is not symplectic (see \cite{ZhMa1988}).  This method is implemented by solving 
	\[
		y_1 = \mathrm{Ad}^\ast_{\exp(-\xi)} (y_0), \quad \text{with} \quad \xi = \frac{h}{2} \left(\nabla H(y_0) + \nabla H(y_1) \right),
	\]
	where $H$ is the point vortex Hamiltonian \eqref{hamiltonian_function}, $y_0, y_1$ are the point vortex locations in $\mathbb{R}^3$, viewed as the dual $\mathfrak{so}(3)^\ast$ of the Lie algebra of the rotation group, and $\mathrm{Ad}^\ast_g(y_0) = gy_0g^{-1}$.
\end{enumerate}

Our conclusion is that of all four methods, the Hopf integrator and the midpoint method on $\mathbb{S}^2$ do a good job of preserving the geometric structure of the point vortex equations. Both methods preserve the vortex moment exactly, while the Hopf integrator exhibits in addition the bounded energy error associated with symplectic integrators. While the Hopf integrator is a little less accurate than the midpoint method for a given step size, it is also somewhat faster, so that both methods perform comparably. The 4th-order Runge--Kutta method and the Lie--Poisson method generally exhibit a linear drift in the conserved quantities.

On the whole, the conservation properties of the midpoint algorithm seem to be somewhat coincidental, and rely on the fact that for the point vortex equations, the algorithm stays on the unit sphere without reprojecting.  It is therefore not clear how to generalize this algorithm to obtain, for instance, higher-order integrators with similar conservation properties.  By contrast, for the variational Hopf integrator it suffices to start from a higher-order version of the discrete Lagrangian \eqref{discrete_lagrangian} to obtain a higher-order variational Hopf integrator.

We implemented our algorithm using various routines from the \textsc{NumPy} and \textsc{SciPy} scientific libraries (see \cite{Ol2007}). A version of our code can be found at 
\url{https://github.com/jvkersch/hopf_vortices}.

\subsection{Stable relative equilibria of vortex rings} \label{sec:polvani_dritschel}

\begin{figure}[h!]
\begin{center}
	\includegraphics[scale=0.55]{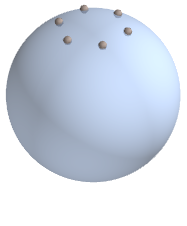}
	\quad
	\includegraphics[scale=0.4]{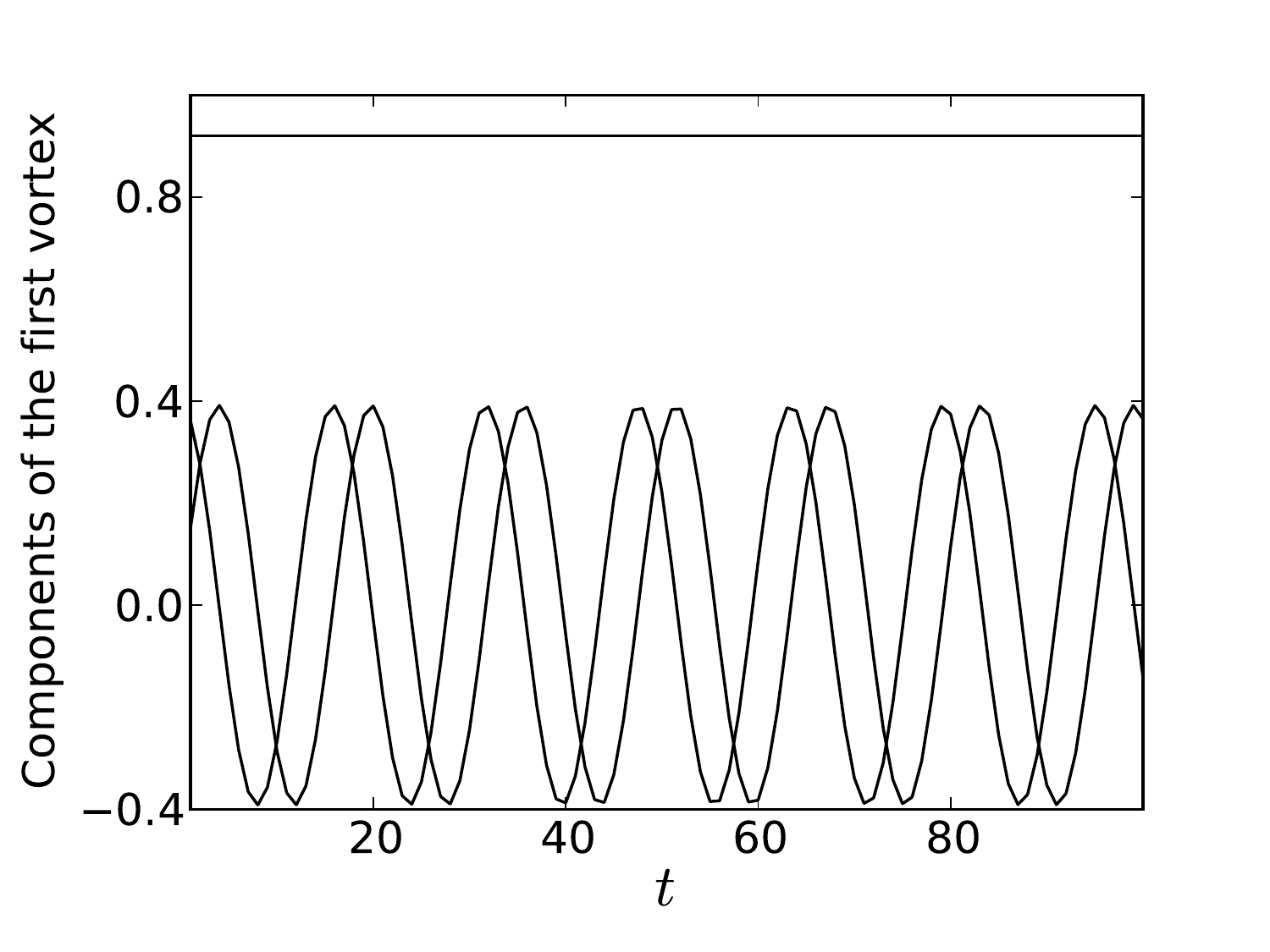}
	\caption{Left: Initial conditions for the 6-vortex Polvani-Dritschel vortex ring.  Right: $x$, $y$ and $z$-component of the first vortex in the Polvani-Dritschel simulation, where the time-step $h = 0.1$.  The trajectory is clearly seen to be periodic. \label{fig:pd-trajectory-variational-short}}
\end{center}
\end{figure}

\cite{PoDr1993} have investigated the behavior of a ring of $N$ equidistant vortices with the same strength $\Gamma$, placed on a circle of fixed latitude on the sphere (see Figure~\ref{fig:pd-trajectory-variational-short}).  They found that this configuration is a stable relative equilibrium, provided that $N \le 7$ and that the colatitude is below a certain critical value (dependent on $N$).  For the case $N = 6$, the critical colatitude is given by $\theta_c = \arccos(2/\sqrt{5}) \approx 0.464$ and the stable relative equilibria satisfy $\theta_0 < \theta_c$.  The vortex ring rotates around the $z$-axis with angular velocity $\Omega = (N-1)\frac{\Gamma}{4 \pi} \frac{z_0}{1 - z_0^2}$, where $z_0 = \cos \theta_0$.

For our simulation, we choose $N = 6$, $\Gamma = 1/6$ and $\theta_0 = 0.40$, so that $\Omega \approx 0.397$ and the period $T \approx 15.819$.   The motion of the first vortex over a number of periods is illustrated in Figure~\ref{fig:pd-trajectory-variational-short}.

\begin{figure}
\begin{center}
	\includegraphics[scale=0.45]{%
		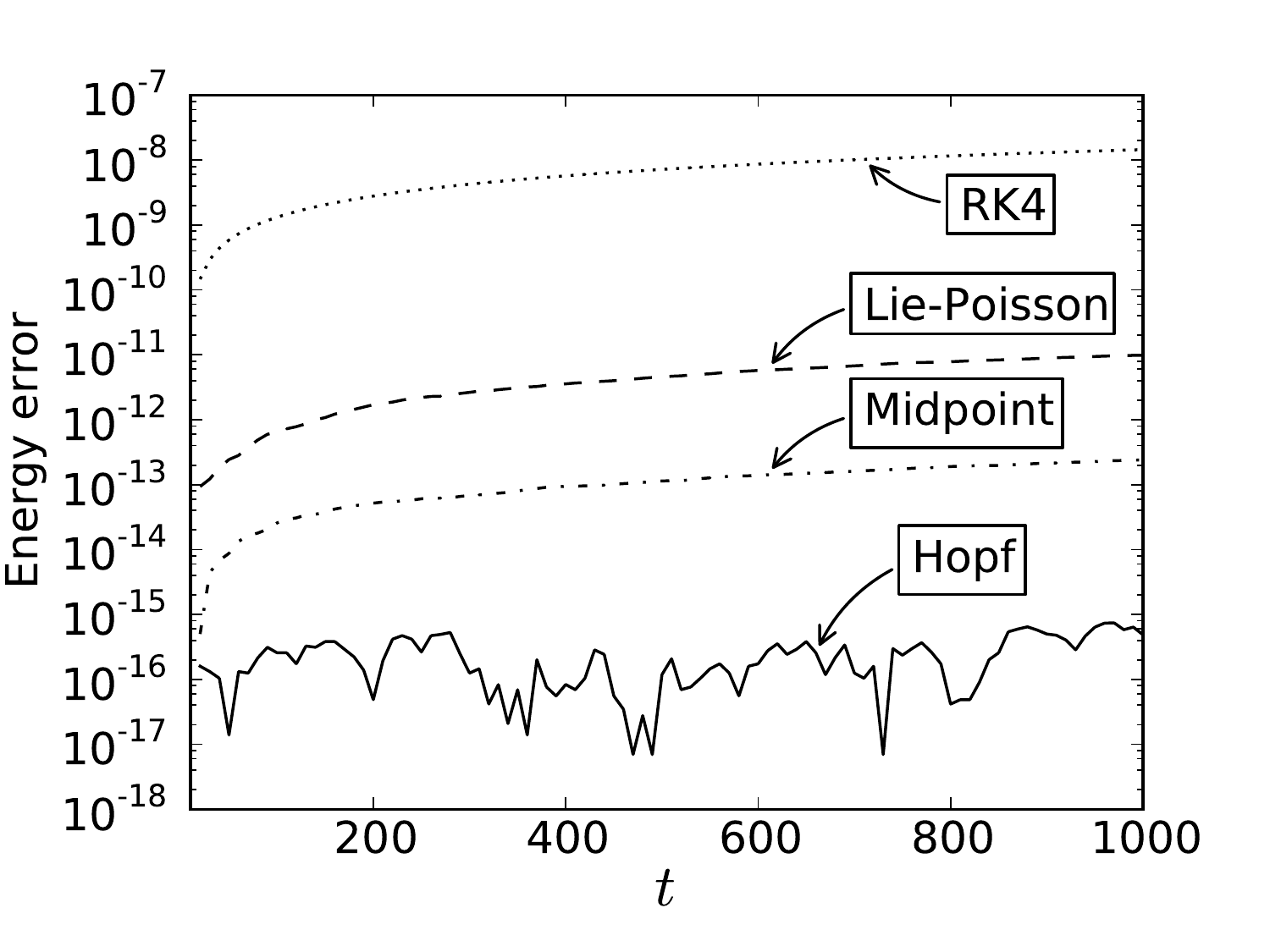}
	\includegraphics[scale=0.45]{%
		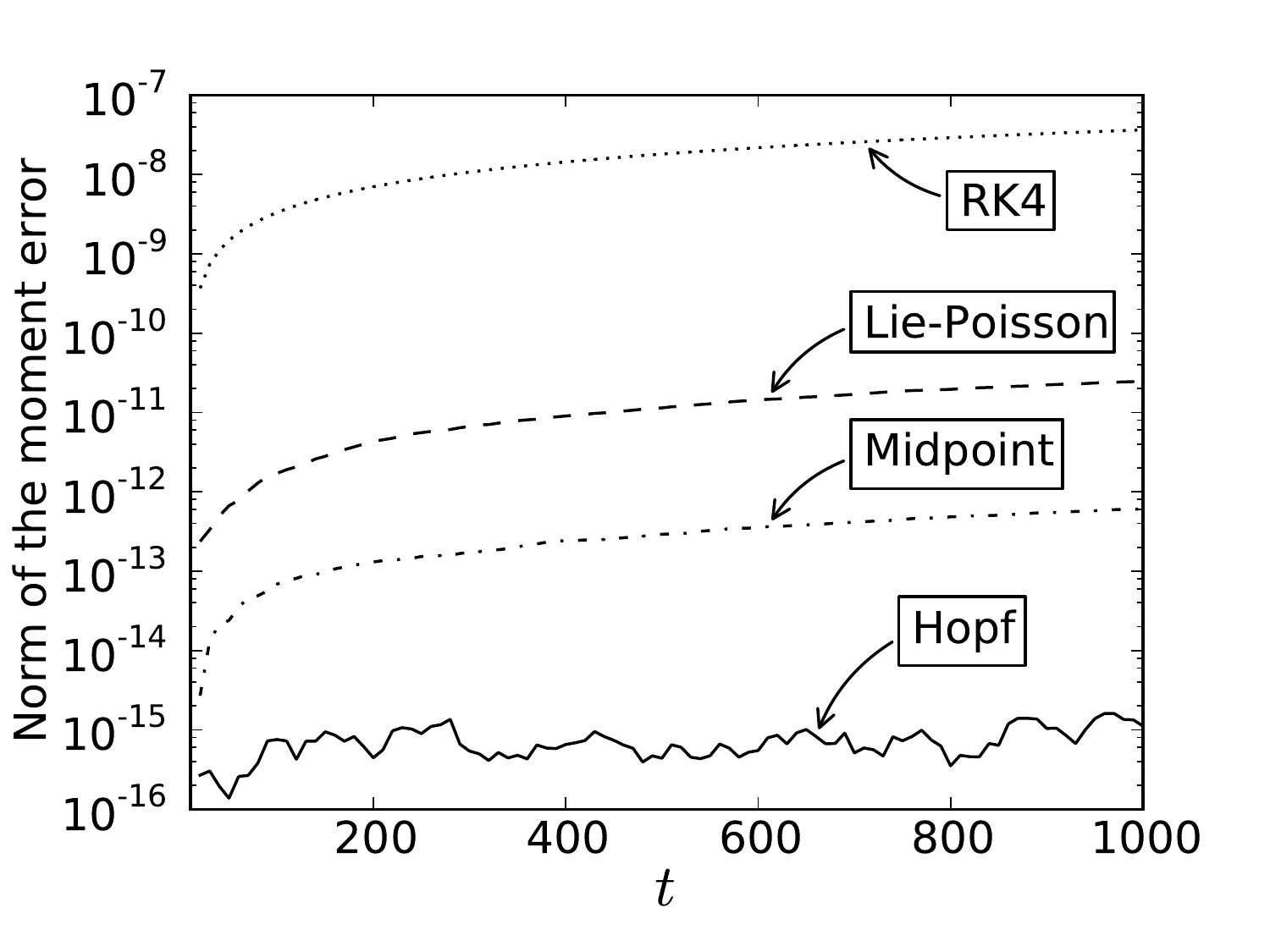}
	\caption{Comparison of the energy and momentum preservation between all four methods for the stable Polvani-Dritschel vortex ring. The Hopf integrator preserves both invariants up to machine precision, while the other integrators exhibit a clear drift.  Here $h = 0.1$ and $\sigma = 0.0$. \label{fig:energy-momentum-comparison}}
\end{center}
\end{figure}

\paragraph{Comparison with other integrators.}  

We next turn to the energy and momentum conservation properties of the numerical integrator.    We simulate the motion of the Polvani-Dritschel vortex ring with time step $h = 0.1$ and regularization parameter $\sigma = 0.0$, for $T = 1000$ units of time, using all four integrators. 

In Figure~\ref{fig:energy-momentum-comparison} we have plotted the absolute energy error $\Delta E := \left| E(t_n) - E(t_0) \right|$ (left) and the moment error $\Delta M := \left\Vert \textbf{M}(t_n) - \textbf{M}(t_0)\right\Vert$ (right) as a function of time.  The Hopf integrator preserves the energy and vortex moment to machine precision, while the other three integrators exhibit drifts in both conserved quantities at various rates.

\paragraph{Numerical order calculation.}  

We know from theoretical considerations that the Hopf integrator is second-order accurate, and so are the two other geometric methods.  We now illustrate this statement by comparing the solution trajectories generated by the Hopf integrator with the exact trajectories.  For 10 choices of time step $h$ between $10^{-4}$ and $10^{-1}$ we run the simulation over $T = 100$ units of time and we compute the absolute error between the numerical and the exact solution. We consider only the first vortex, since the trajectories of the other vortices differ from the first by a rigid rotation.  More precisely, for each integrator we do the following: if $\mathbf{x}_{\mathrm{exact}}(t_n)$ is the exact position of the first vortex at time $t_n = nh$ and $\mathbf{x}_{\mathrm{int}}^{n, h}$ is the numerical trajectory, then we compute 
\[
	\Delta_h := \max_{n} \left\Vert 
		\mathbf{x}_{\mathrm{exact}}(t_n) - \mathbf{x}_{\mathrm{int}}^{n, h} \right\Vert 
\]
for each of the selected time steps. For the sake of comparison, we have also included the simulation results for the 2nd-order Heun's method composed with projection onto $\mathbb{S}^2$, which is labeled on the figure as RK2.

\begin{figure}[h!]
\begin{center}
	\includegraphics[scale=.45]{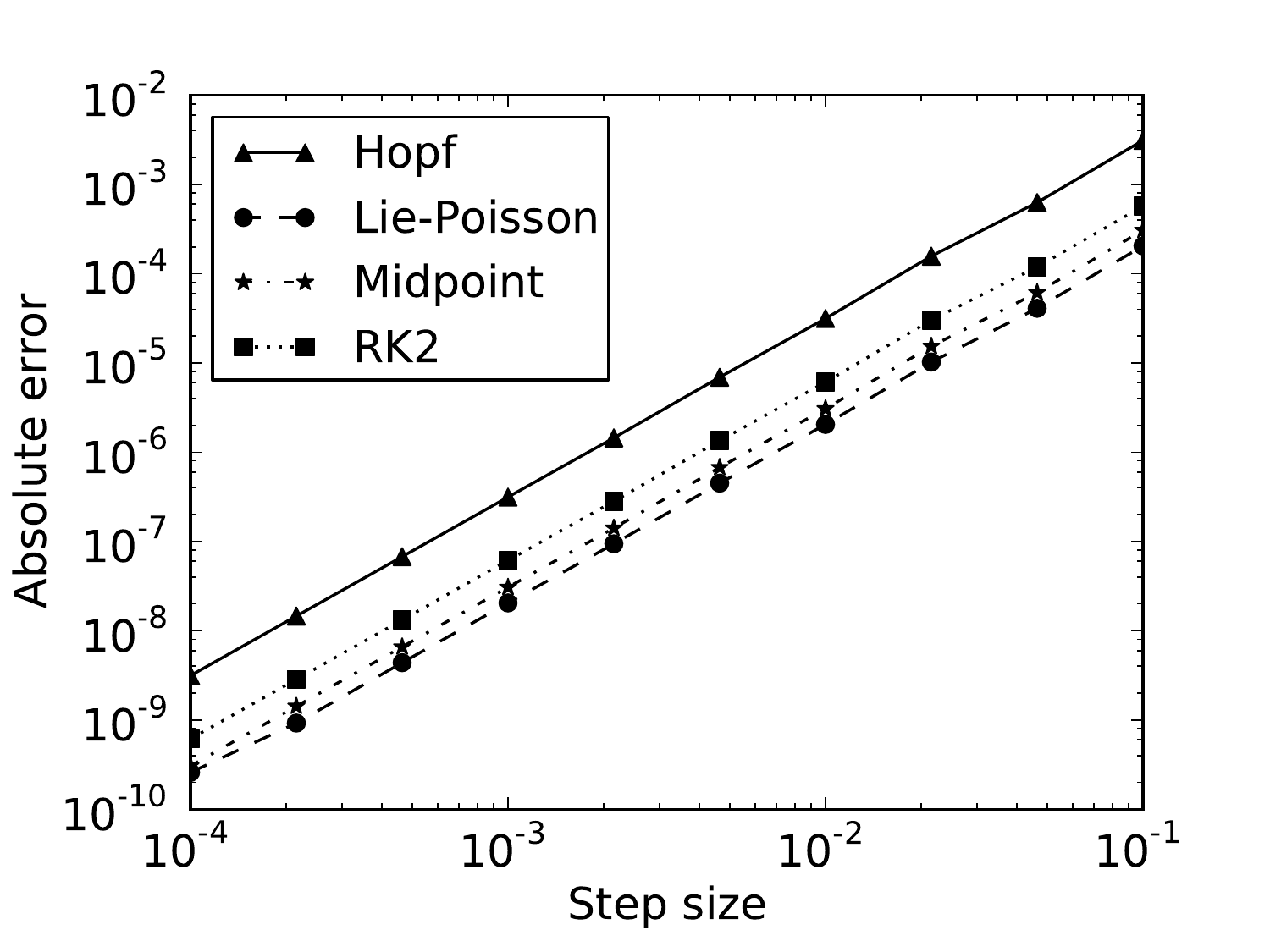}
	\includegraphics[scale=.45]{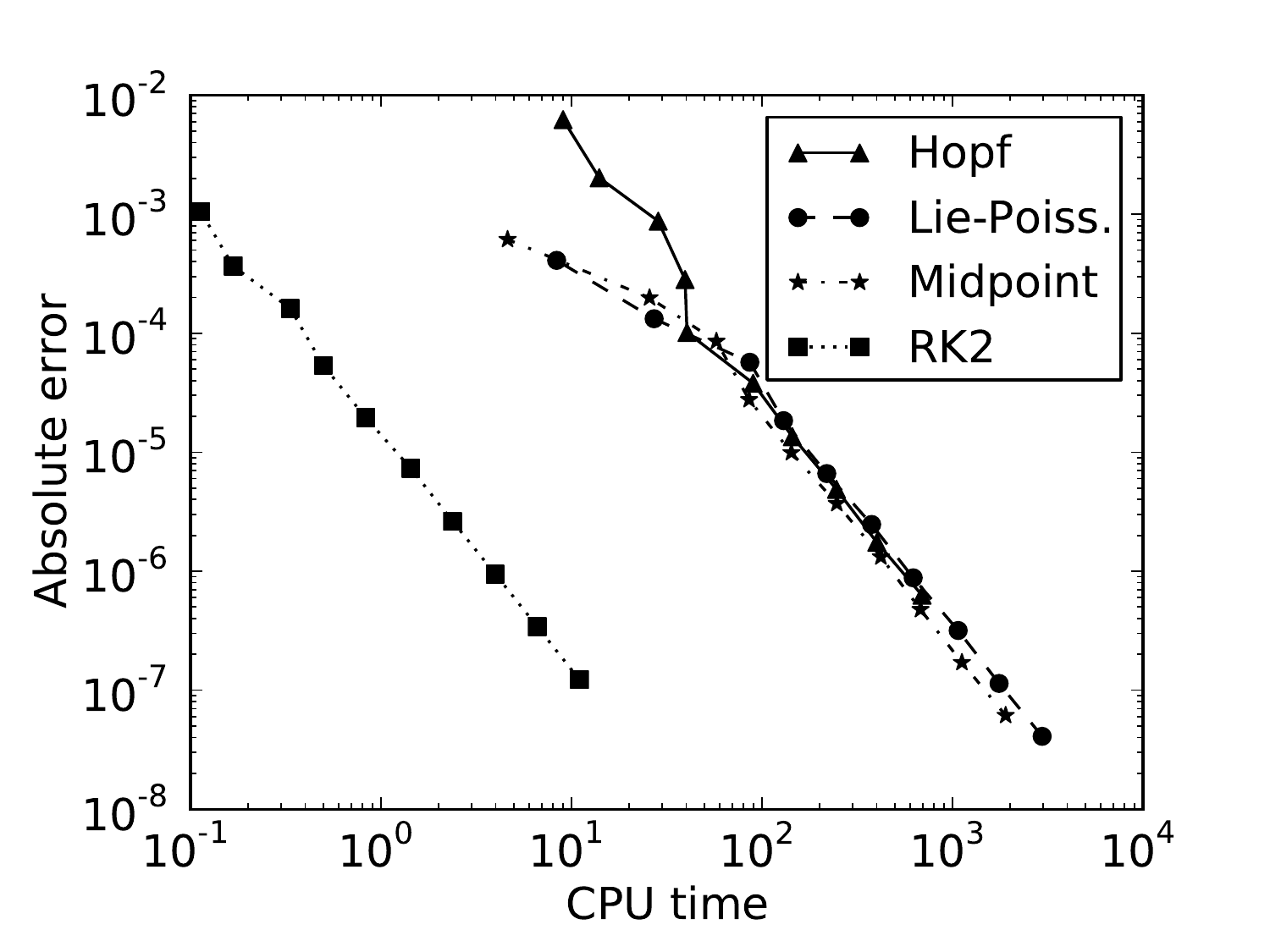}
	\caption{Left: Absolute error for each of the four integrators for the Polvani-Dritschel vortex ring over $T = 100$ units of  time. All four integrators are second-order accurate in time.  Right: Absolute error as a function of CPU time expended, again for $T = 100$ units of time. All three geometric integrators exhibit very similar behavior in accuracy vs. computational cost, and Runge--Kutta is much cheaper than the geometric integrators for the same accuracy.\label{fig:variational-order}}
\end{center}
\end{figure}

Figure~\ref{fig:variational-order} (left) shows a plot of absolute errors versus time steps for the three geometric integrators as well as RK2.  All four integrators are of second-order.  On the right pane of Figure~\ref{fig:variational-order}, we have plotted the obtained accuracy for each of the  methods as a function of the expended CPU time.

We see that, apart from a transient regime for large step sizes in which the Hopf integrator is an order of magnitude slower, all three geometric methods perform comparably. In relative terms, RK2 clearly outperforms all three geometric methods, since with modest computational expense many orders of accuracy are obtained.  This is partly a result of the fact that the Polvani-Dritschel vortex ring is a relatively simple, periodic vortex system. We will see in the examples below that for non-equilibrium configurations, the energy and vortex moment slowly drift from their true values when integrated with a non-symplectic integrator.

\subsection{The spherical von K\'arm\'an vortex street}

\begin{figure}[h!]
\begin{center}
	\subfloat[][]{\includegraphics[scale=.45]{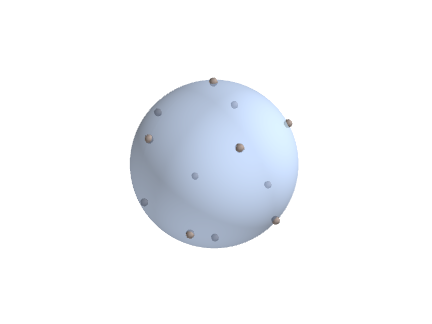}
		\label{fig:svs_ic}}
	\subfloat[][]{\includegraphics[scale=.45]%
		{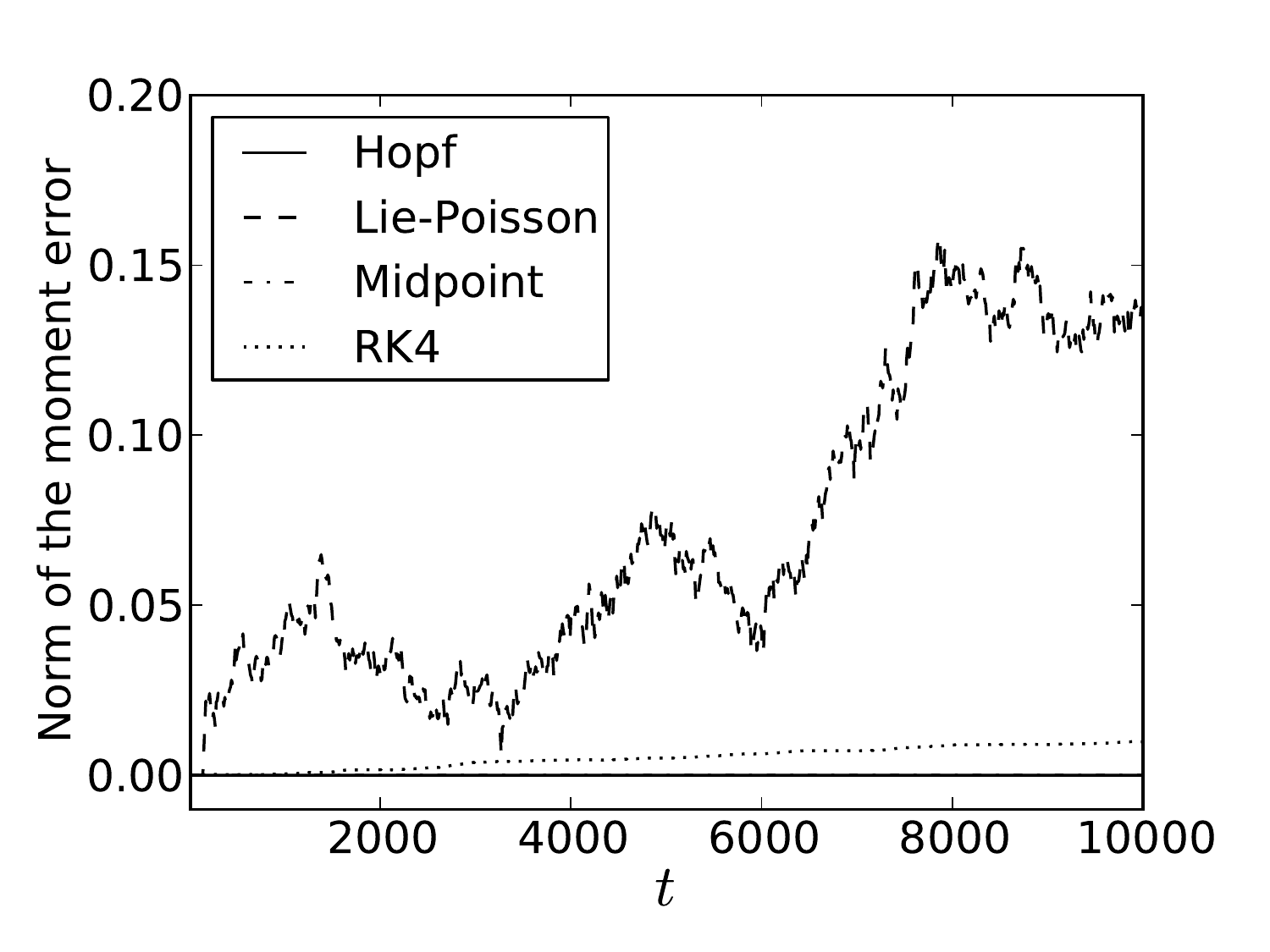} \label{fig:svs_mom}} \\
	\subfloat[][]{\includegraphics[scale=.45]%
		{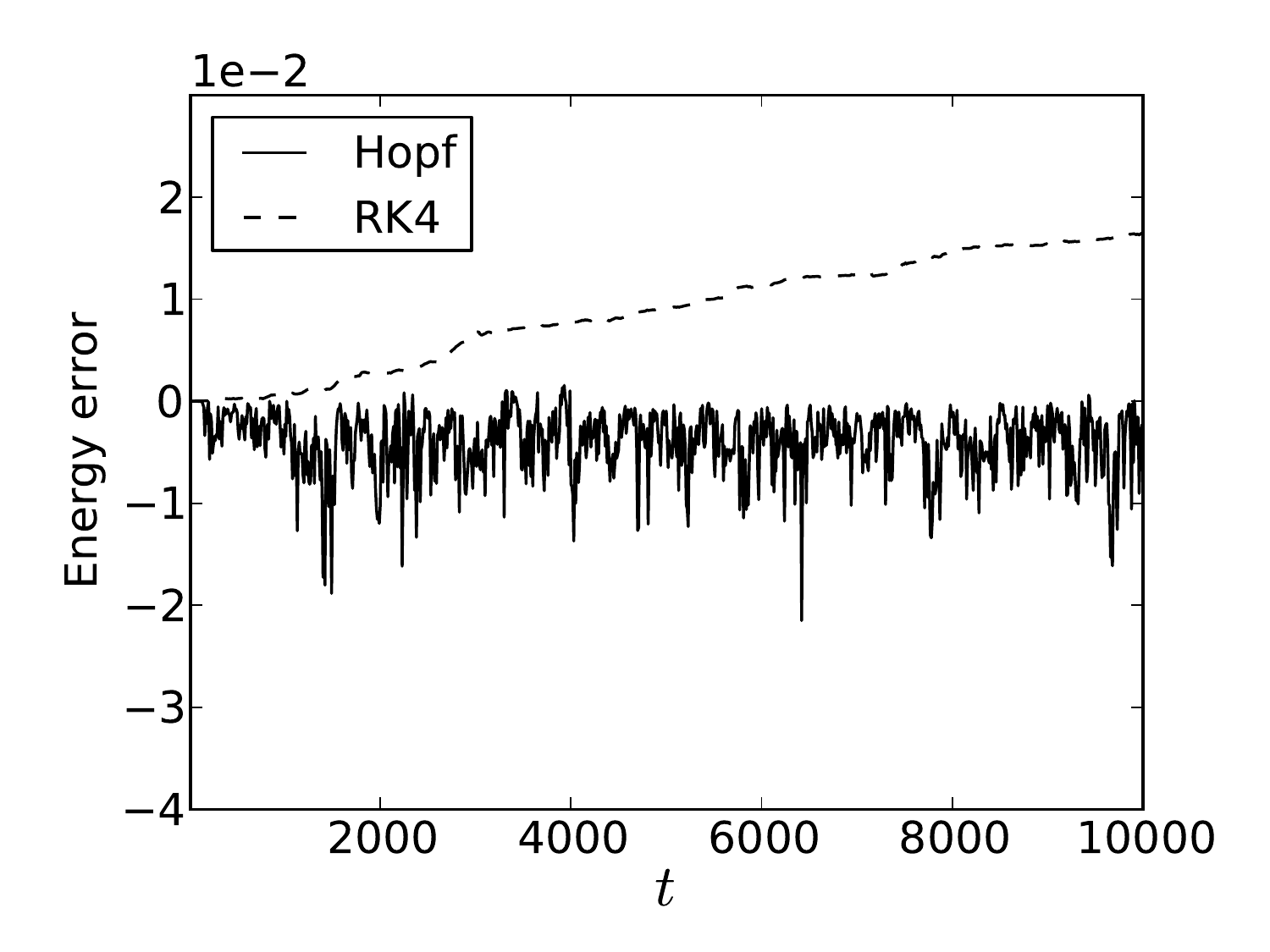} \label{fig:svs_en1}}
	\subfloat[][]{\includegraphics[scale=.45]%
		{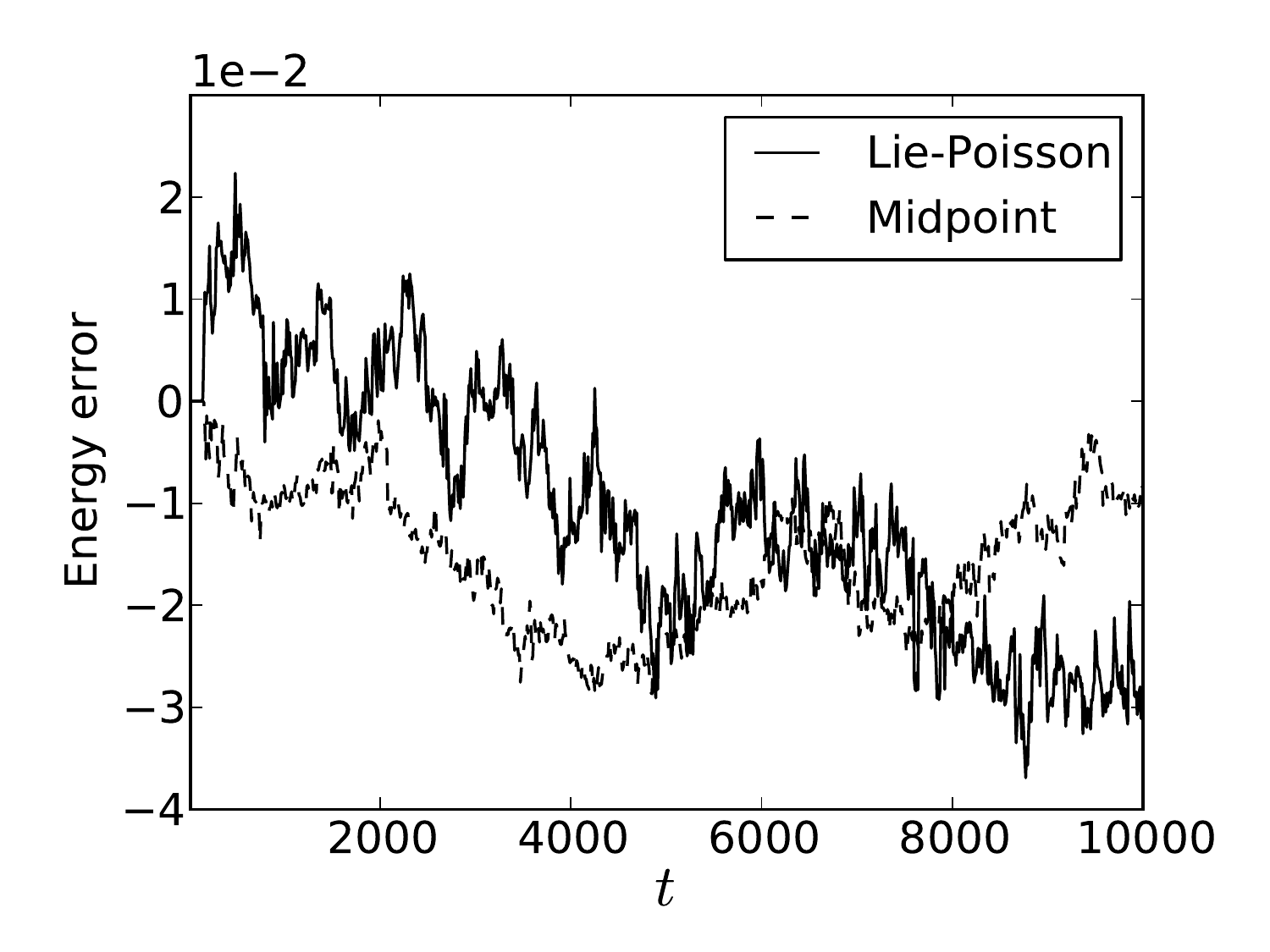} \label{fig:svs_en2}}
	\caption{Long term simulation of the spherical von K\'arm\'an vortex street. (a) Initial configuration of the vortices. (b) Long-term conservation of the moment of vorticity. The Hopf and midpoint integrators preserve the moment exactly while RK4 and the Lie--Poisson method exhibit a linear drift. (c) and (d) Long-term energy preservation of the Hopf and RK4 integrator (left) and the Lie--Poisson and midpoint method (right).  Of the four integrators, the Hopf integrator is the only one to exhibit bounded energy oscillations over long integration times.}
\end{center}
\end{figure}

An important class of relative equilibria consists of the single and double von K\'arm\'an vortex streets on the sphere described by \cite{ChKaNe2009}.  The single vortex street consists of two staggered arrays of vortices, each consisting of $N$ equidistant vortices of strength $\Gamma$, at fixed colatitudes $\phi = \phi_1$ and $\phi = \pi - \phi_1$, together with vortices of strength $\Gamma_n$ and $\Gamma_s$ at the north and the south pole, respectively (see Figure~\ref{fig:svs_ic}).

For the simulations in this section, we take the number of vortices in each ring to be $N = 5$, and we set the colatitude equal to $\phi_1 = \pi/3$.  The vortex strength for the ring vortices is set equal to unity, $\Gamma = 1$, while the polar vortices satisfy $\Gamma_n = - \Gamma_s = 1/2$.  This configuration forms a relative equilibrium which rotates around the $z$-axis with period $T = 10.85$.  Based on the behavior of the planar Von K\'arm\'an vortex street, it is believed that this relative equilibrium is unstable, although no rigorous stability analysis exists, to the best of our knowledge.

In the simulation, the equilibrium becomes unstable and breaks up after a short amount of time, leading to aperiodic motion of the vortices.  In this regime, the energy is not  exactly preserved by the Hopf integrator, but exhibits bounded oscillations, as is to be expected from a symplectic integrator.   For this simulation, we used time step $h = 0.5$ and regularization parameter $\sigma = 0.25$ and we ran the simulation for $10\, 000$ time units. 

%

In Figure~\ref{fig:svs_mom}, we have plotted the error in the vortex moment. By construction, the Hopf integrator and the midpoint method on $\mathbb{S}^2$ are exactly moment-preserving, and their error in the moment is seen to vanish, whereas the moment error for the Lie--Poisson and the RK4 method grows linearly. In Figures~\ref{fig:svs_en1} and \ref{fig:svs_en2} we have plotted  the energy error for each of the integrators. Of the four integrators, the Hopf integrator is the only one that exhibits bounded error in the energy, with energy oscillations of the order of $10^{-2}$. The energy errors for the RK4 method and the Lie--Poisson method grow secularly, while the energy for the midpoint method on $\mathbb{S}^2$ resembles a random walk.

\subsection{Self-similar collapse of three vortices}

\begin{figure}[t]
\begin{center}
	\includegraphics[scale=.45]{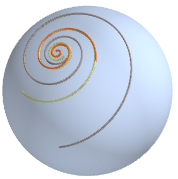}
	\caption{Trajectories of 3 colliding vortices, for the initial conditions described in the text. \label{fig:colliding3}}
\end{center}
\end{figure}

It is well known that certain configurations of point vortices on the sphere will collapse to a point in finite time.  For 3 vortices, necessary and sufficient conditions for collapse were given by \cite{KiNe1998} while \cite{Sa2008} identified an open set of initial conditions for collapse of 4 vortices.  We focus here on the case of 3 vortices.

We simulate the motion of three vortices with strengths $\Gamma_1 = \Gamma_2 = 1$, $\Gamma_3 = -1/2$ placed at the vertices of a triangle with side lengths $l_{12} = \sqrt{3}/2$, $l_{23} = \sqrt{2}/2$ and $l_{31} = 1$.    For this configuration, it can be calculated that collapse occurs after $\tau_- \equiv 4\pi(\sqrt{23} - \sqrt{17}) \cong 8.4537$ units of time.  The trajectories of the colliding vortices are shown in Figure~\ref{fig:colliding3}.  Note that these initial conditions are for the unregularized system, i.e. \eqref{vortexequations} with $\sigma = 0$.  Adding some regularization to the system effectively amounts to imposing a minimum distance on the vortices and will prevent the vortex configuration from collapsing to a single point.

\begin{figure}[ht!]
\begin{center}
	\includegraphics[scale=.45]{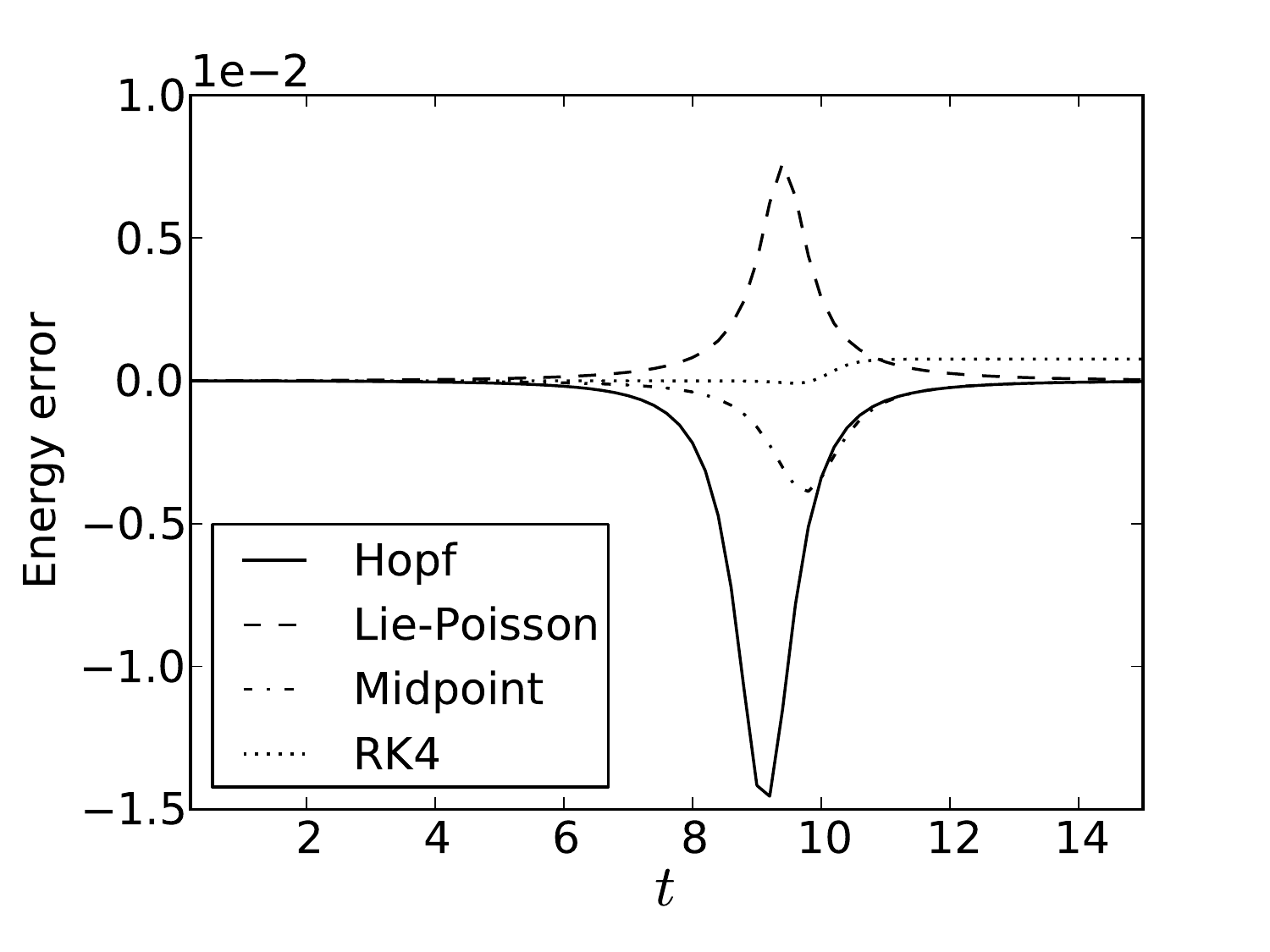}
	\includegraphics[scale=.45]{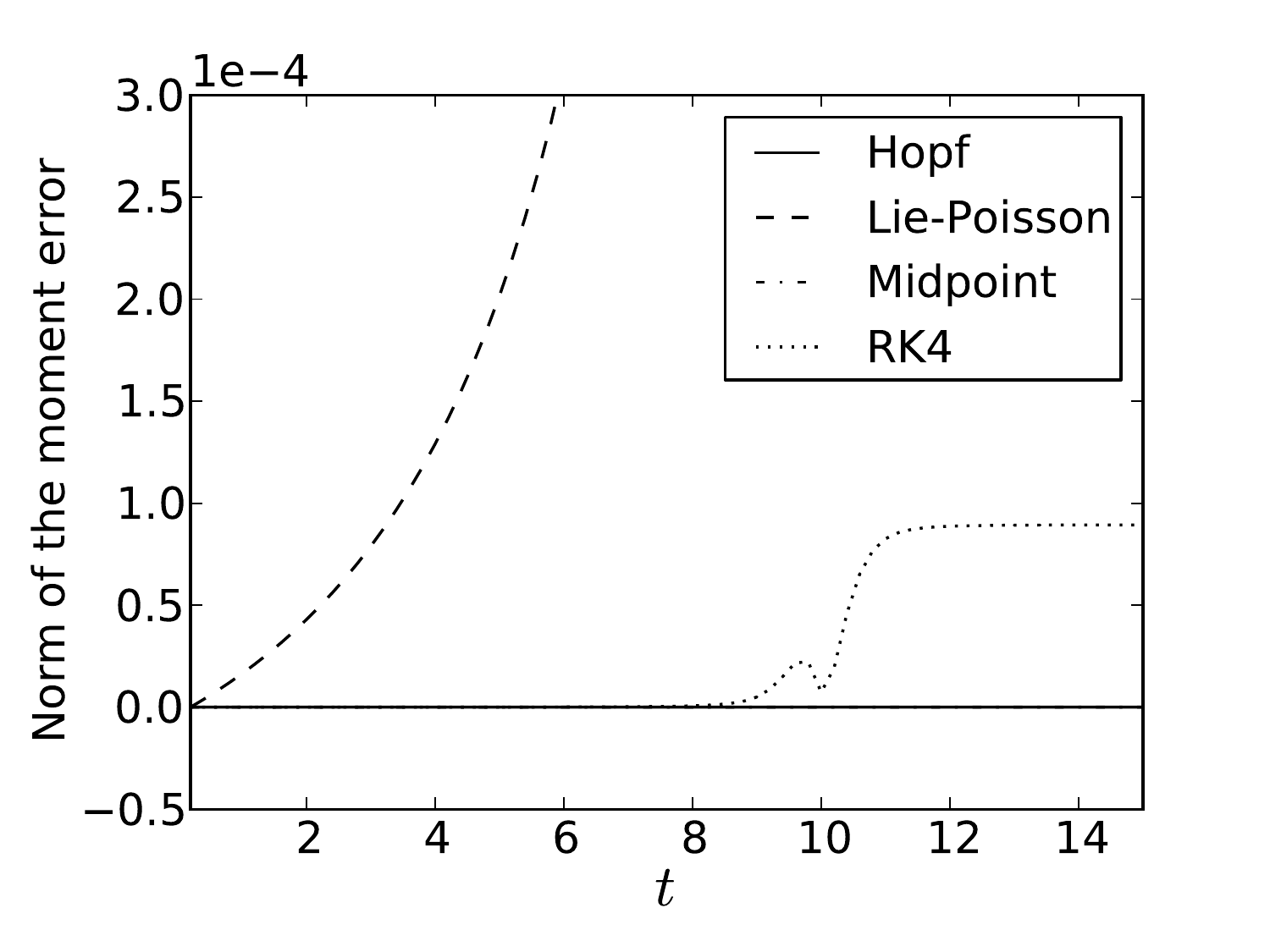}
	\caption{Energy (left) and moment (right) conservation for the geometric  integrators and 4th-order Runge--Kutta close to vortex collapse, which happens for the unregularized system at $t \approx 8.45$. While RK4 conserves the energy and moment better than the geometric integrator up to the collapse, the energy and moment settle down in a different value after collapse.  Here, $h = 0.1$ and $\sigma = 0.1$. \label{fig:collapse_nosigma}}
	\includegraphics[scale=.45]{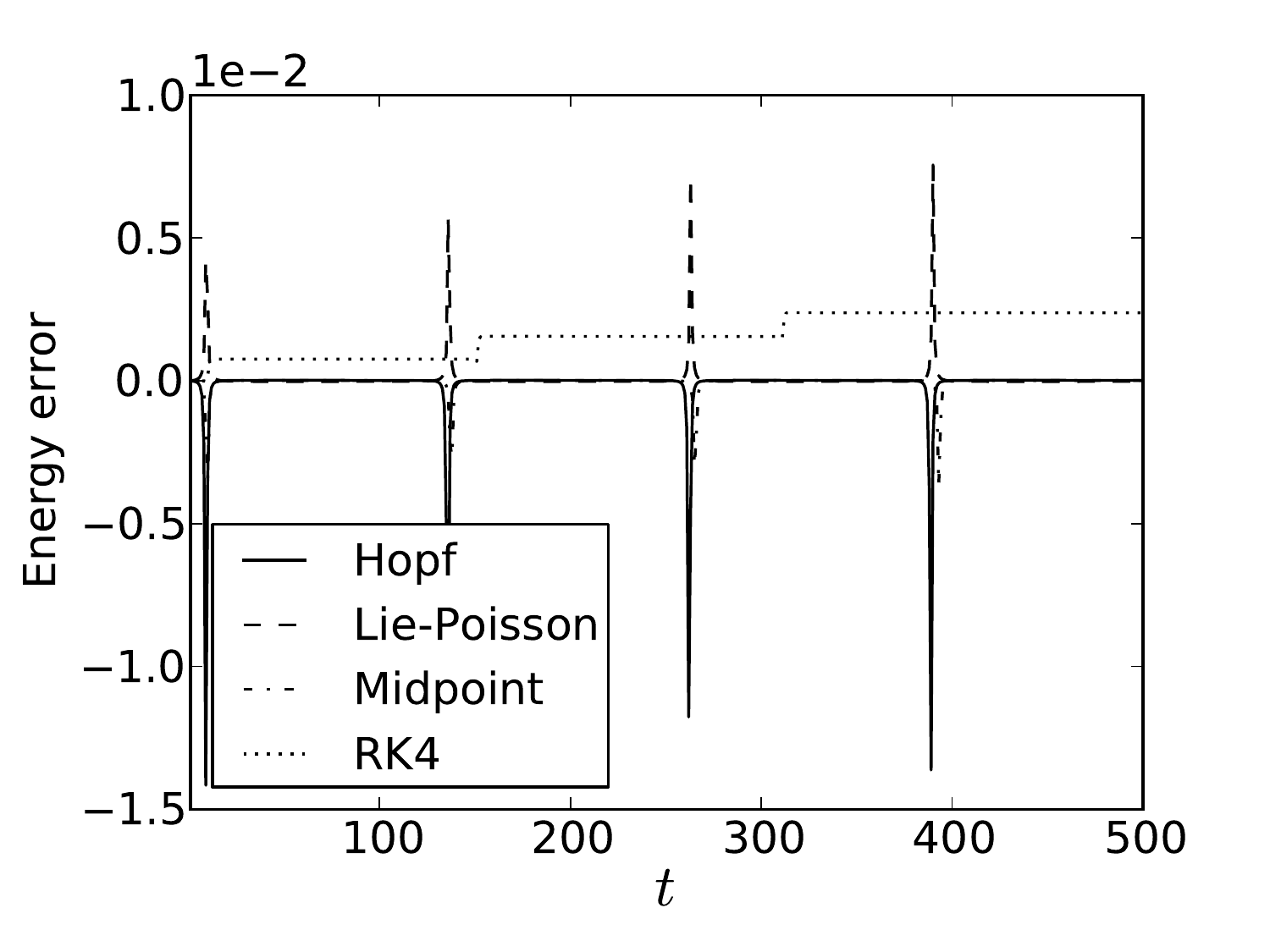}
	\includegraphics[scale=.45]{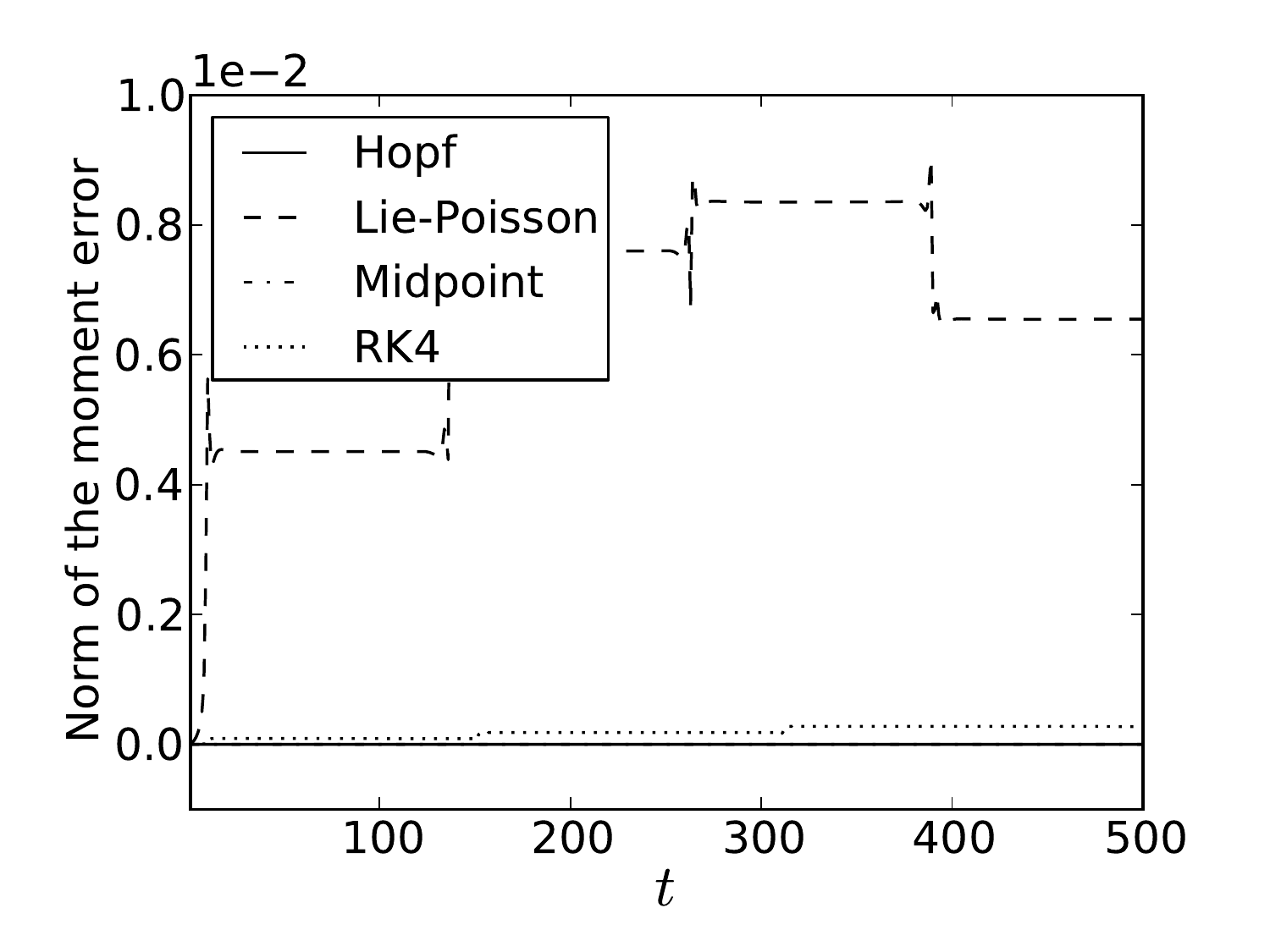}
	\caption{%
	Numerical simulation of colliding point vortices for $T = 500$ units of time, where $h = 0.1$ and $\sigma = 0.10$.  At regular instances of time, there are collapse events, indicated by the spikes in the figures. Whereas the Hopf and the midpoint method preserve energy and the vortex moment reasonably well away from collapse events, the energy increases with each collapse for RK4, and the moment increases with each collapse for both the Lie--Poisson and RK4 method.
	 \label{fig:long_collapse}}
\end{center}
\end{figure}

We simulate the system first with a moderate time step, $h = 0.1$, and some amount of regularization, $\sigma = 0.10$, for $T = 15$ units of time.  On Figure~\ref{fig:collapse_nosigma} we see that the geometric integrators, including the Hopf integrator, perform a better job of preserving the energy and vortex moment than Runge--Kutta: while all integrators show some buildup in the energy and moment error around the time of the collapse, the energy and moment return to their original values after the collapse for the geometric integrators, but settle down at a slightly different value for Runge--Kutta.  

For long-term simulations, this effect is more pronounced.  After the near-collapse, the three vortices travel past each other and (nearly) collapse again at a later time, a situation which repeats itself periodically afterwards.   Figure~\ref{fig:long_collapse} shows that for every collapse event, the Runge--Kutta simulation incurs a jump in the energy and the moment, whereas the geometric integrators manage to preserve these invariants without any noticeable secular trend.  Note also that the period of time between two collapse events increases gradually for Runge--Kutta, but stays (roughly) constant for the geometric integrators. This, in particular, is an indication of the fact that the energy drift exhibited by RK4 changes the qualitative nature of the dynamics.

\subsection{Large ensembles of vortices} \label{sec:large_ensembles}

For our last example, we return to the numerical simulation of vortex rings on the sphere, as in Section~\ref{sec:polvani_dritschel}.  We put 40 vortices of strength $\Gamma = 1/8$ at equal distances from each other on a circle with colatitude $\theta_0 = \arccos(0.9)$. This configuration approximates a vortex sheet on the sphere, and exhibits the typical Kelvin-Helmholtz instability associated with vortex sheets.

\begin{figure}
\begin{center}
	\includegraphics[scale=.5]{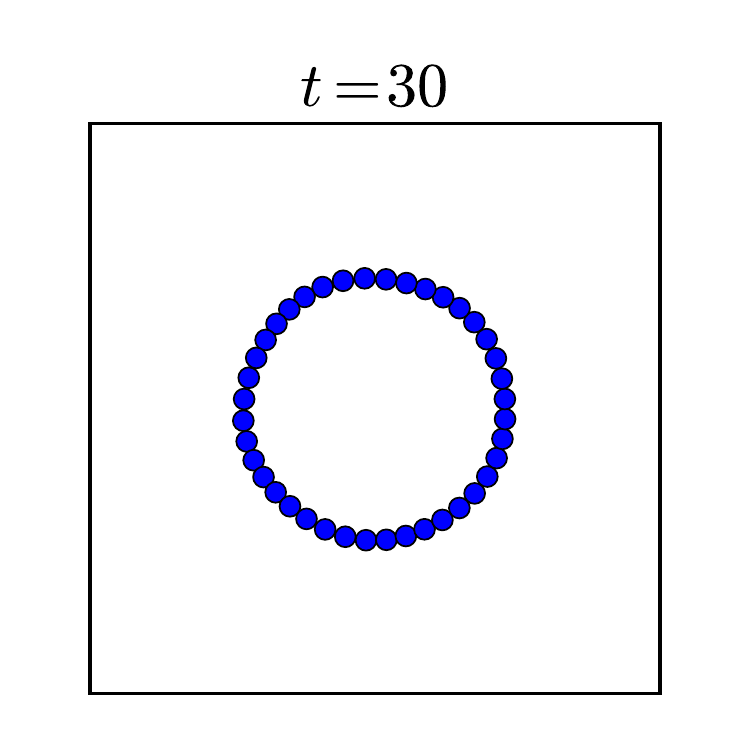}
	\includegraphics[scale=.5]{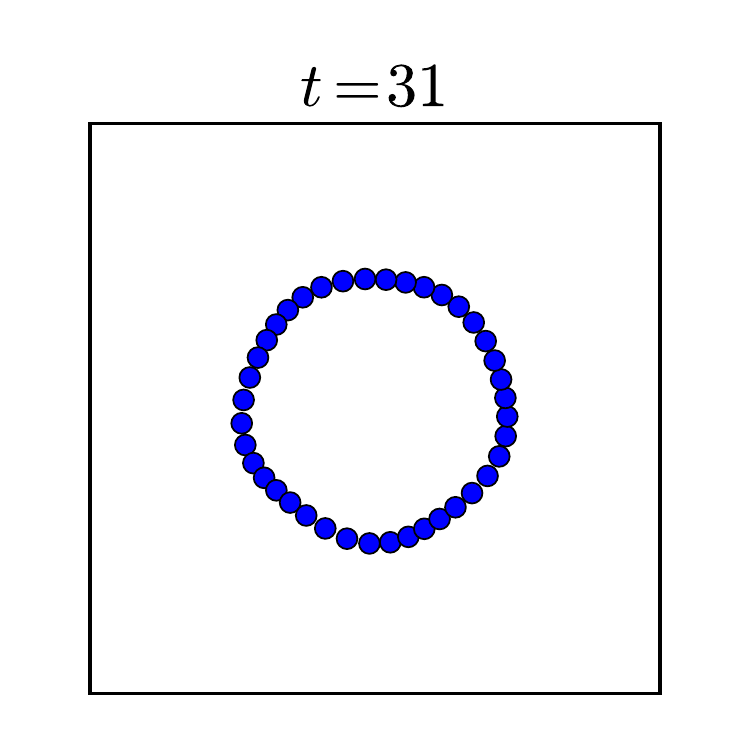}
	\includegraphics[scale=.5]{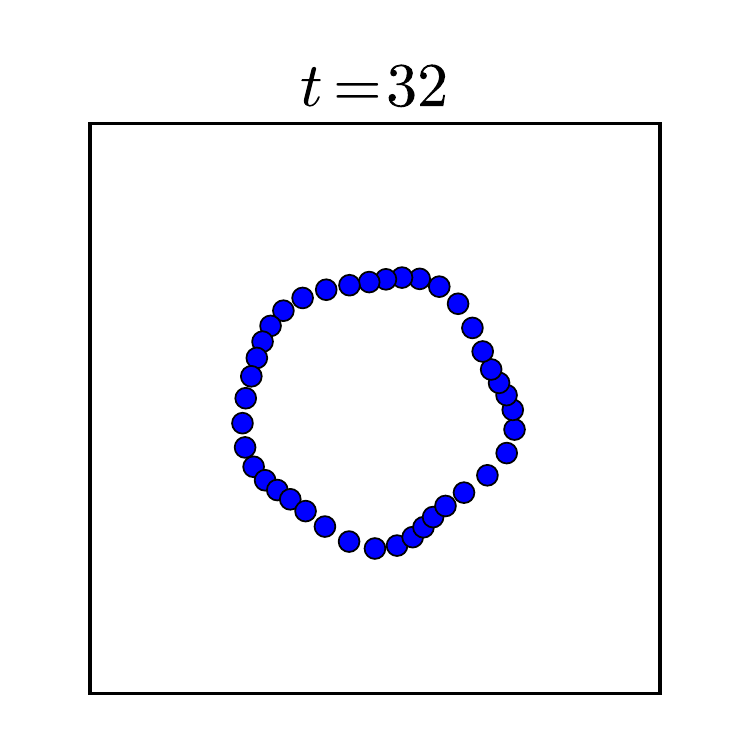} \\
	\includegraphics[scale=.5]{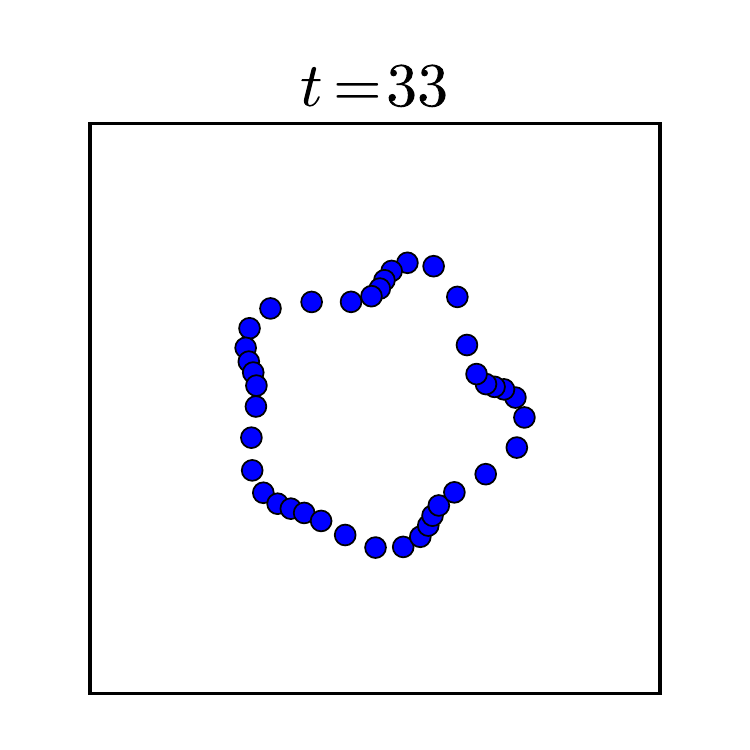}
	\includegraphics[scale=.5]{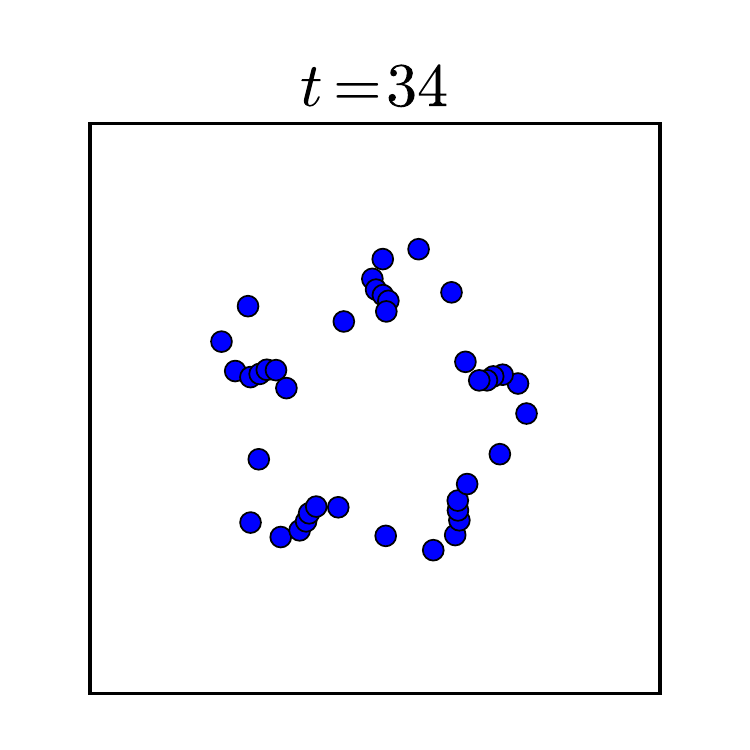}
	\includegraphics[scale=.5]{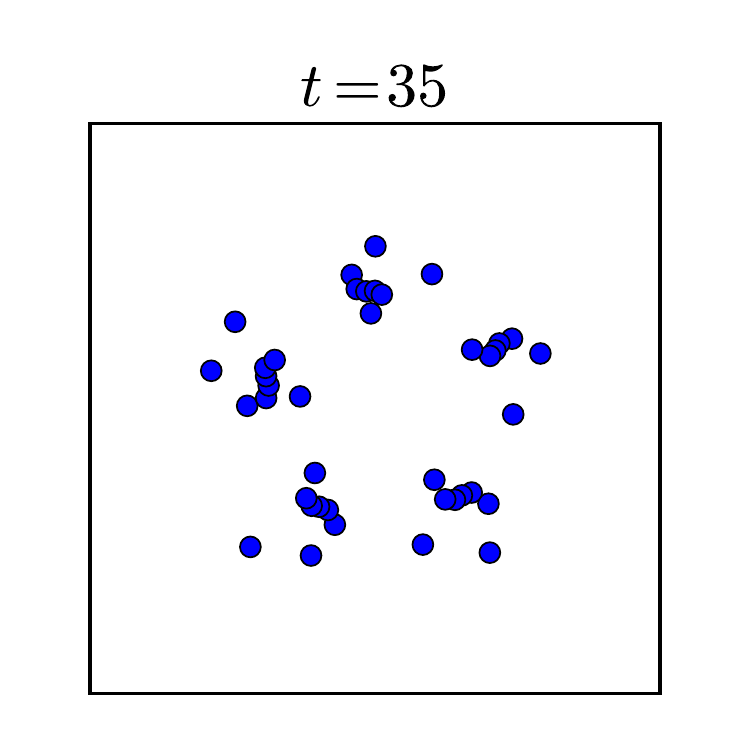}
	\caption{Kelvin-Helmholtz instability of a vortex sheet on the sphere, approximated by $40$ point vortices of strength $1/8$.  A sequence of snapshots is shown around the onset of instability. For the purpose of visualization, the sphere has been projected onto the plane by means of stereographic projection from the North pole, but the simulation is done directly on the sphere. \label{fig:vortex_frames}}
\end{center}
\end{figure}

\begin{figure}
\begin{center}
	\includegraphics[scale=0.45]{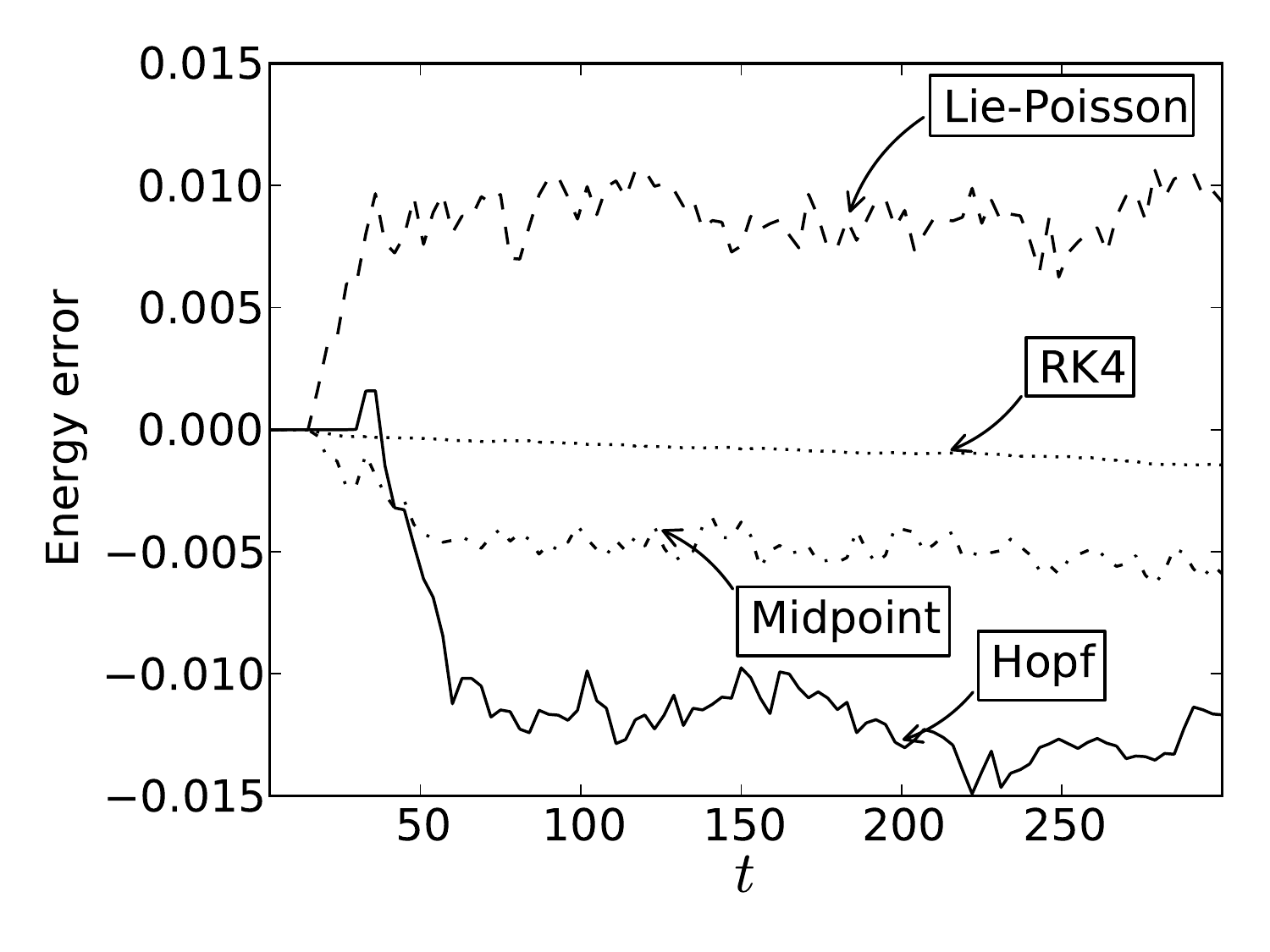}
	\includegraphics[scale=0.45]{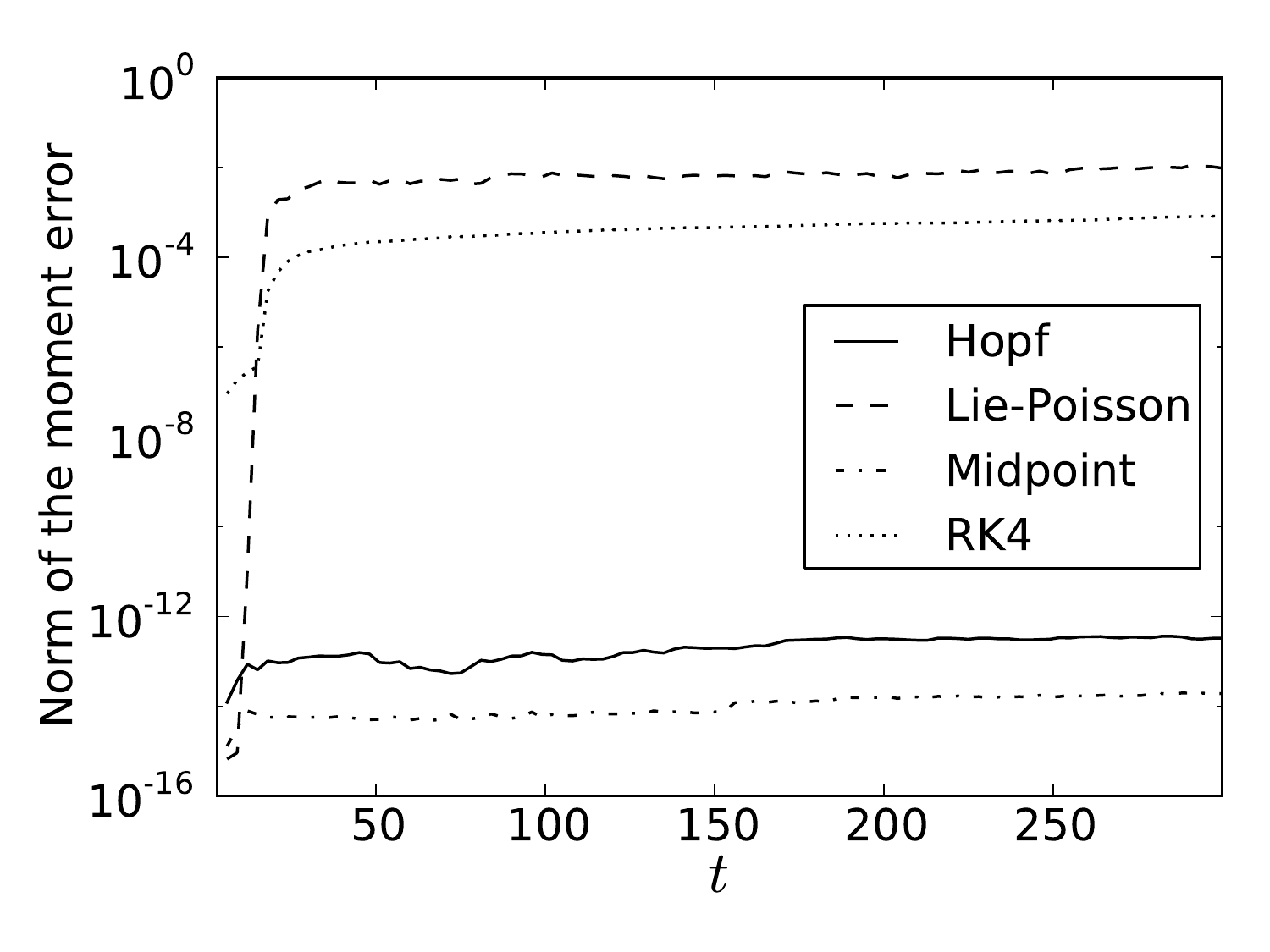}
	\caption{Plot of the energy error (left) and the moment error (right) for the vortex sheet approximated by 40 point vortices. The onset of instability for Hopf integrator is around $t = 30$, while for the other integrators instability sets in much earlier, around $t = 15$. After the breakup of the relative equilibrium, the vortices move in a non-equilibrium manner, as witnessed by the (bounded) error in the energy.\label{fig:big-ring-em}}
\end{center}
\end{figure}

We simulate this system with a fairly large timestep, $h = 0.1$, and some amount of regularization, $\sigma = 0.1$.  In Figure~\ref{fig:big-ring-em} we have plotted the energy and moment error for moderate integration times.  While the Hopf integrator does not preserve the energy any better than the other geometric methods, it preserves the vortex ring somewhat longer than the other integrators: whereas for the non-symplectic integrators instability sets in around $t = 15$, the Hopf integrator preserves  stability until $t = 30$.

Around $t = 32$, we see on Figure~\ref{fig:vortex_frames} that the vortex ring deforms into a pentagonal configuration, which then curls up around $t = 34$ and breaks up for $t \ge 35$. For a theoretical interpretation of the Kelvin-Helmholtz instability on the sphere, see~\cite{Sakajo2004}.

\section{Conclusions and outlook}

In this paper, we have used the Hopf fibration to construct a linear Lagrangian on the three-sphere $\mathbb{S}^3$, whose Euler--Lagrange equations project down to the point vortex equations on $\mathbb{S}^2$.  In the second part of the paper, we have used this Lagrangian formulation to construct a variational integrator for point vortices on the sphere.  Below, we discuss some possibilities for future research.

\paragraph{Extension to higher-order integrators.} Our Lagrangian approach to the construction of discrete point vortex integrators can be extended without major difficulties to the construction of integrators whose numerical order is higher than two.  It suffices to take a discretization in \eqref{discrete_lagrangian} which is of higher than second-order.  The standard theory of Lagrangian variational integrators (see \cite{MaWe2001}) then ensures that the resulting discrete equations of motion will have the same order of accuracy as the discrete Lagrangian.

\paragraph{Other Lie group methods.}  We have constructed a Lie group variational integrator by directly discretizing the linear Lagrangian \eqref{linear_lagrangian}.  Another approach is due to \cite{BoMa2009} (see also \cite{KoMa2011}).  Here the Lagrangian is first written in a left-trivialization of $TSU(2) \cong SU(2) \times \mathfrak{su}(2)$ by mapping $(g, \dot{g})$ to $(g, \xi)$, where $g^{-1} \dot{g} = \xi$, and this equation is then discretized and added to the variational principle using a Lagrange multiplier, giving rise to the so-called \emph{Hamilton-Pontryagin variational principle}.  A similar variational principle, known as the \emph{Clebsch variational principle}, was pioneered in \cite{CoHo2009}.  It would be of considerable interest to discretize our linear Lagrangian \eqref{linear_lagrangian} using these augmented variational principles and to compare the properties of the resulting discrete mechanical system with our straightforward discretization.

\paragraph{Statistical mechanics of large numbers of point vortices.}  The statistical theory of vortex motion set forth in \cite{On1949} predicts that, under certain energetic conditions, like-signed vortices will tend to cluster over time.  We have not attempted to extract any statistical information from the simulation of large number of vortices on the sphere, but it would be interesting to do so.  To alleviate the $O(N^2)$-cost of computing the point vortex Hamiltonian while maintaining the symplectic nature of the integrator, a geometric fast multipole method like the one developed in \cite{ChDaFa2010} could be used.

\paragraph{Vorticity distributions on the sphere and other surfaces.}  Point vortices represent the simplest non-trivial distributions of vortices on the sphere.   The methods proposed in this paper are expected to generalize without any significant difficulty to the case of vortex blobs or patches of vorticity on the sphere (see \cite{Ch1973, Ne2001}).  

To treat vortical distributions on other surfaces, the following construction from prequantization could be used.  Recall that the Hopf fibration is the fiber bundle associated to the quantum line bundle on $\mathbb{S}^2$ associated with the area form; see e.g. \cite{Wo1992}.  For the motion of point vortices on a surface $\Sigma$ with integral area form, one can follow a similar route and lift the motion of the vortices to (the principal fiber bundle associated to) the quantum line bundle, for which a similar relation as \eqref{pullback_quantization} will continue to hold.

Moreover, PDEs such as the KdV and nonlinear Schr\"odinger equation can also be formulated using a linear Lagrangian, and we hope that the methods introduced in this paper may be useful for these systems as well.

\subsection*{Acknowledgments}

We are very grateful to the referees of this paper, whose comments and observations significantly improved our exposition.

We would like to dedicate this paper to the memory of Hassan Aref, whose kind encouragement and insightful remarks at the 2010 SIAM-SEAS meeting at the University of North Carolina, Charlotte, provided the initial stimulus for this work.  Furthermore, we would like to thank J. D. Brown, C. Burnett, B. Cheng, F. Gay-Balmaz, M. Gotay, D. Holm, E. Kanso, S. D. Kelly, P. Newton,  T. Ohsawa, B. Shashikanth and A. Stern for stimulating discussions and helpful remarks.

M. L. and J. V. are partially supported by NSF grants DMS-1010687, CMMI-1029445, and DMS-1065972. J. V. is on leave from a Postdoctoral Fellowship of the Research Foundation--Flanders (FWO-Vlaanderen).  This work is supported by the \textsc{irses} project \textsc{geomech} (nr. 246981) within the 7th European Community Framework Programme.

\appendix
\section{Analysis of a planar vortex integrator} \label{sec:planar}

In this appendix, we show that the integrator of \cite{RoMa2002} for point vortices in the plane shares a number of remarkable features with the Hopf integrator, which stems from the fact that both systems are derivable from a linear Lagrangian.   Similar observations, but for the numerical integration of canonical Hamiltonian systems, were made by \cite{Brown2006}.

\paragraph{Decomposition into one-step methods.}

 \cite{RoMa2002} start from the linear Lagrangian \eqref{plane_lagrangian}, which they discretize by setting
\[
	L_d(z_0, z_1) = h L \left( (1-\alpha) z_0 + \alpha z_1, 
		\frac{z_1 - z_0}{h}\right),
\]
where $\alpha \in [0, 1]$ is a real interpolation parameter.  The equations of motion derived from this Lagrangian are given by 
\begin{equation} \label{planar_pv_equations}
	\frac{z_{n+2} - z_n}{2h} = \alpha f(z_{n+\alpha})
		+ (1-\alpha) f(z_{n+1+\alpha}),
\end{equation}
where $z_{n+\alpha} := (1-\alpha)z_n + \alpha z_{n+1}$ and $f(z)$ is the right-hand side of the vortex equations \eqref{plane_equations}.  It turns out that for $\alpha = 1/2$, they can be written as the composition of a one-step method and its adjoint.  To see this, we specialize to the case $\alpha = 1/2$ and use the fact that the original Lagrangian $L$ is linear in the velocities to write 
\[
	L_d(z_0, z_1) =  L( z_{1/2}, z_1) - L( z_{1/2}, z_0),
\]
and we define $L_{d, +}(z_0, z_1, h) := L( z_{1/2}, z_1)$ and $L_{d, -}(z_0, z_1, h) := - L( z_{1/2}, z_0)$, so that $L_d = L_{d, +} + L_{d, -}$.  Consider the \emph{adjoint} $L_d^*$ of a discrete Lagrangian $L_d$, which is defined by $L_d^\ast(z_0, z_1, h) := - L_d(z_1, z_0, -h)$ (see \cite{MaWe2001}). Then, we have that 
\[
	L_{d, +}^\ast(z_0, z_1, h) = L_{d, -}(z_0, z_1, h), 
\]
and vice versa. This definition is motivated by the fact that the adjoint of the discrete Euler--Lagrange flow of a discrete Lagrangian is given by the discrete Euler--Lagrange flow of the adjoint discrete Lagrangian.

The composition of the discrete Euler--Lagrange flow of two discrete Lagrangians is given  by the discrete Euler--Lagrange flow of a composition discrete Lagrangian that is the sum of the two original discrete Lagrangians. As a result, the discrete Euler--Lagrange flow for $L_d$ is given by the composition of the discrete Euler--Lagrange flows for $L_{d, +}$ and its adjoint $L_{d,+}^*=L_{d, -}$. These discrete flows can be viewed as one-step methods, and are typically only first-order accurate, but their composition is symmetric and therefore has even order of accuracy, and is, in particular, second-order accurate. 

Lastly, we remark that for the point-vortex Lagrangian \eqref{plane_lagrangian} the discrete Lagrangians $L_{d, +}$ and $L_{d, -}$ \emph{coincide}, which means that each of them is individually self-adjoint. As a result, the underlying one-step method is second-order.  In fact, it can easily be seen that for $\alpha = 1/2$, the point vortex equations \eqref{planar_pv_equations} can be written as the composition of the implicit midpoint method 
\[
	\frac{z_{n+1}-z_n}{h} = f(z_{n + 1/2}) 
\]
with itself.  This method is clearly second-order accurate.

For the case of point vortices on the sphere the Lagrangians  $L_{d, +}$ and $L_{d, -}$ still coincide, but in order to recover the equations of motion \eqref{discrete_EL} and to enforce the constraint $\left< \varphi^{n+1}, \varphi^{n+1}\right> = 1$, different constraint forces have to added to the discrete flow.   As a result, the underlying one-step methods, which are the maps $\Phi_h$ and $\Psi_h$ defined at the end of Section~\ref{sec:projected_midpoint}, no longer coincide and are individually only first-order accurate (unless the underlying Hamiltonian is $\mathbb{S}^1$-invariant), although their composition is second-order accurate.

\begin{figure}
\begin{center}
	\includegraphics[scale=.45]{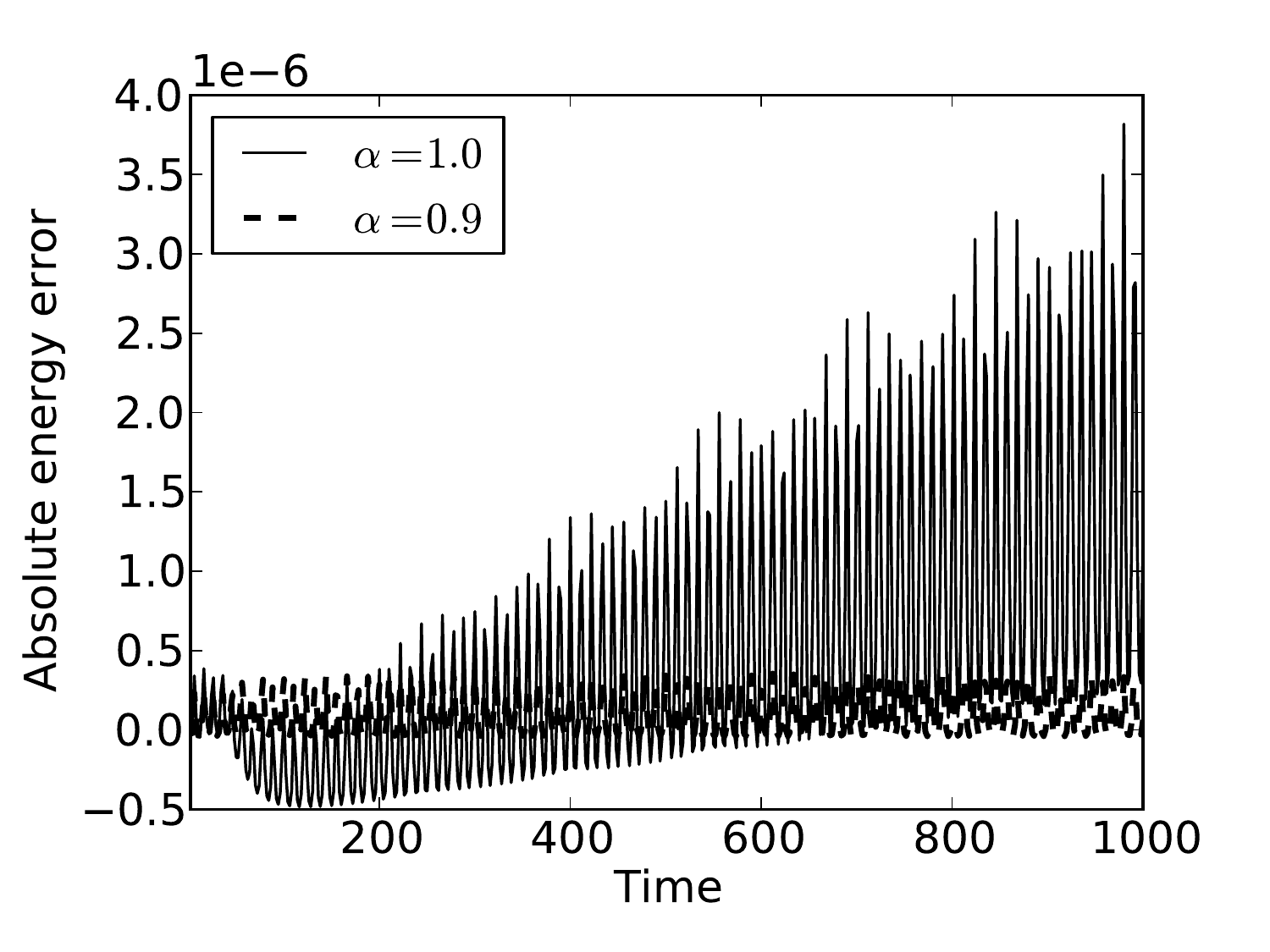}
	\caption{For the four-vortex problem described in the text, the energy error exhibits a linear drift for the integrator with $\alpha = 1$ (solid line) but stays bounded whenever $\alpha \ne 1$; here $\alpha = 0.9$ is shown (dashed line). \label{fig:linear-growth}}
\end{center}
\end{figure}

\paragraph{The choice $\alpha = 0,1$ for the interpolation parameter.}  The method \eqref{planar_pv_equations} is implicit for all choices of $\alpha$ except $\alpha = 0, 1$, in which case the equations become 
\begin{equation} \label{unstable_midpoint}
	\frac{z_{n+2} - z_n}{2h} =  f(z_{n+1}).
\end{equation}
This method turns out to be the \emph{symmetric explicit midpoint method} (see \cite{HaLuWa2002}), which is well-known to exhibit parasitic oscillatory solutions.  These solutions can easily be observed in the dynamics of point vortices: in Figure~\ref{fig:linear-growth}, we have plotted the energy error for a simulation of a four-vortex problem with vortex strengths $\Gamma = (.1, .3, -.2, -.4)$ and initial conditions $z = (0, .5i, 1, .7 + .6i)$.  For the simulation where $\alpha = 0.9$ the energy error is bounded, while for the simulation employing $\alpha = 1.0$ there is a clear linear drift in the energy error.  The time step used for both simulations was $h = 0.1$.

This is in clear contrast to the construction of variational integrators for \emph{nondegenerate} Lagrangians, for which any choice of interpolation parameter $\alpha$ will result in a stable, second-order integrator.  

Similar instabilities exist for the case of point vortices on the sphere: the discrete equations \eqref{sphere_eq_unstable}, for instance, exhibit the same instabilities as \eqref{unstable_midpoint}, despite being variational.

\section{A variational integrator for non-$\mathbb{S}^1$-invariant Hamiltonians} \label{app:no_invariance}

In Section~\ref{sec:projected_midpoint} we were able to obtain the implicit midpoint version \eqref{implicit_midpoint_method_S3} of the Hopf integrator on $\mathbb{S}^3$ based on the assumption that the Hamiltonian $H$ is invariant under the action of $(\mathbb{S}^1)^N$ on $(\mathbb{S}^3)^N$. When the Hamiltonian is not invariant, this simplification is no longer possible, and the equations \eqref{projected_discrete2} and \eqref{LHS} must be solved directly. In this Appendix, we outline a strategy for doing so, based on the geometry of the group $SU(2)$.

\paragraph{Implementing the unit-length constraint: the Cayley map.}

Given initial conditions $(\varphi^{n-1}, \varphi^n)$, we first compute the slack variables $d^n_\alpha$ using \eqref{LHS}. We must now solve \eqref{projected_discrete2} for $\varphi^{n+1}$, and we need to impose the unit-length constraint \eqref{disc_unit_length}.  This can be done conveniently using the geometry of $SU(2)$: we write the update map $\varphi^n \mapsto \varphi^{n+1}$ as  
\begin{equation} \label{cayley_update}
	\varphi^{n+1} = U^{n} \varphi^n,
\end{equation}
where $U^{n}$ is an element of $SU(2)$.   This ensures that the length of $\varphi^n$ stays constant over time, since
\[
	(\varphi^{n+1})^\dagger \varphi^{n+1}  = 
	(\varphi^n)^\dagger (U^{n} )^\dagger U^{n} \varphi^n = 
	(\varphi^n)^\dagger \varphi^n,
\]
so that, in particular, $\left\Vert \varphi^n \right \Vert = 1$ implies that $\left\Vert \varphi^{n+1} \right \Vert = 1$.

The equations \eqref{projected_discrete2} for $\varphi^{n+1}$ can now be expressed as 
\begin{equation} \label{projected_discrete3}
\Re \left[ (\varphi^n)^\dagger (\mathrm{i} \sigma_\alpha) \left(-\mathrm{i} \Gamma ( U^{n} - I_{2\times2}){\varphi}^{n}  
		 + \frac{h}{2} D_{{\varphi}^\dagger}H(\varphi^{n+1/2})\right)  \right] = - d^n_\alpha,
\end{equation}
where $\varphi^{n+1/2}$ in the Hamiltonian can be expressed in terms of $U^{n}$ and $\varphi^{n}$ by
\[
   \varphi^{n+1/2} = \frac{1}{2} \left( \varphi^{n} + \varphi^{n+1} \right) = 
     \frac{1}{2} ( I + U^{n}) \varphi^n.
\]

These equations can be solved for $U^{n}$ directly, but a computationally more advantageous approach is as follows.  As long as the step size $h$ is small, the update matrix $U^{n}$ will be in a neighborhood of the identity element in $SU(2)$.  We now parametrize that neighborhood by means of the \emph{Cayley transform}
$\mathrm{Cay}: \mathfrak{su}(2) \to SU(2)$, given by 
\begin{align*}
	\mathrm{Cay}(A) & = (I + A)(I - A)^{-1}.  
\end{align*} 
That is, we search for an element $A^{n} \in \mathfrak{su}(2)$ such that $U^{n} = \mathrm{Cay}(A^{n})$ will solve \eqref{projected_discrete3}.  The advantage is that $\mathfrak{su}(2)$ is a linear space, and that no constraints need to be imposed on $A^{n}$, as the range of the Cayley map is automatically contained within $SU(2)$.  A standard nonlinear solver can therefore be used to find $A^{n}$. This is analogous to the approach used in \cite{LeLeMc2009} to implement the unit-length constraint on $S^2$ by updating the solution on the sphere using a $SO(3)$ action that is parametrized by the Cayley transform from $\mathfrak{so}(3)$ to $SO(3)$.

\paragraph{Computational savings.}
Significant computational savings can be obtained by rewriting the Cayley map in a more convenient form.   We recall from \eqref{vectorrep} that $\mathfrak{su}(2)$ is isomorphic with $\mathbb{R}^3$, and we denote the vector representation of $A^{n}$ by $\mathbf{a}^{n} \in \mathbb{R}^3$.  A small calculation then shows that the Cayley transform can be expressed as 
\begin{equation} \label{explicit_cayley}
	U^{n} = \mathrm{Cay}(A^{n}) = \frac{1}{1 + \left\Vert \mathbf{a}^{n} \right\Vert^2} 
		\left( (1 - \Vert \mathbf{a}^{n} \Vert^2) I + 2 A^{n} \right),
\end{equation}
so that 
\[
	U^{n} - I_{2\times2} = \frac{2}{1 + \Vert \mathbf{a}^{n} \Vert^2} 
		\left(   A^{n} - \Vert \mathbf{a}^{n} \Vert^2 I_{2\times2} \right).
\]

The terms proportional to $\Gamma$ in \eqref{projected_discrete3} can then be written as 
\begin{align}
 \Re \left[\Gamma (\varphi^n)^\dagger \sigma_\alpha  ( U^{n} - I_{2\times2}){\varphi}^{n} \right] 
 	& =\frac{2\Gamma}{1 + \left\Vert \mathbf{a}^{n} \right\Vert^2}
	\Re \left[ (\varphi^n)^\dagger \sigma_\alpha  ( A^{n} - \Vert \mathbf{a}^{n} \Vert^2 I_{2\times2} ){\varphi}^{n} \right] 
		\nonumber \\
	& = \frac{2\Gamma}{1 + \left\Vert \mathbf{a}^{n} \right\Vert^2}
	\left( 
		\Re \left[(\varphi^n)^\dagger \sigma_\alpha A^{n} \varphi^n \right]  
		- \Vert \mathbf{a}^{n} \Vert^2 \Re \left[(\varphi^n)^\dagger \sigma_\alpha \varphi^n \right]
	 \right) \nonumber \\
	 & = \frac{- 2\Gamma}{1 + \left\Vert \mathbf{a}^{n} \right\Vert^2} ( \mathbf{a}^{n} \times \mathbf{x}^{n} + 
	 	\Vert \mathbf{a}^{n} \Vert^2 \mathbf{x}^{n} )_\alpha, \label{rewritten_gamma_terms}
\end{align}
where we have used the expression \eqref{hopf_pauli} for the Hopf fibration to write $x^n_\alpha = (\varphi^n)^\dagger \sigma^\alpha \varphi^n$, as well as the identity
\begin{align*}
	(\varphi^n)^\dagger \sigma_\alpha A^{n} \varphi^n & = 
	\mathrm{i}\sum_{\beta=1}^3 (\mathbf{a}^{n})_\beta  (\varphi^n)^\dagger \sigma_\alpha \sigma_\beta \varphi^n \\
	& = \mathrm{i} (\mathbf{a}^{n})_\alpha - (\mathbf{a}^{n} \times \mathbf{x}^{n})_\alpha,
\end{align*}
which follows easily from \eqref{pauli_mult}.  

Similarly, the terms in \eqref{projected_discrete3} involving the derivatives of the Hamiltonian can be written using \eqref{der_relation} as 
\begin{multline*}
	\Re \left( (\varphi^n)^\dagger 
		(\mathrm{i} \sigma_\alpha) D_{{\varphi}^\dagger}H(\varphi^{n+1/2}) \right)
	= \\ \frac{1}{1 + \left\Vert \mathbf{a}_n \right\Vert^2}
	\left( \mathbf{x}^n \times \nabla {H}^{n+1/2}_{\mathbb{S}^2} - 
		(\mathbf{a}^n \cdot \mathbf{x}^n) \nabla {H}^{n+1/2}_{\mathbb{S}^2}
		- (\mathbf{a}^n \times \mathbf{x}^n) \times \nabla{H}^{n+1/2}_{\mathbb{S}^2}
	\right)_\alpha,
\end{multline*}
where $H_{\mathbb{S}^2}$ is the original point vortex Hamiltonian \eqref{hamiltonian_function}.

Combining this expression with \eqref{rewritten_gamma_terms}, we get that the first-order equations \eqref{projected_discrete3} for $U^{n}$ are equivalent to the following nonlinear equation for $\mathbf{a}^{n}$:
\begin{multline} \label{nonlinear_a}
	  -2\Gamma 
	  ( \mathbf{a}^{n} \times \mathbf{x}^{n} + \Vert \mathbf{a}^{n} \Vert^2 \mathbf{x}^{n} ) 
	  +  \frac{h}{2} 
	\Big(  \mathbf{x}^{n} \times \nabla {H}^{n+1/2}_{\mathbb{S}^2} 
		 - 
		(\mathbf{a}^{n} \cdot \mathbf{x}^{n}) \nabla {H}^{n+1/2}_{\mathbb{S}^2} 	 \\
		- (\mathbf{a}^{n} \times \mathbf{x}^{n}) \times \nabla{H}^{n+1/2}_{\mathbb{S}^2}\Big)
		= -(1 + \left\Vert \mathbf{a}^n \right\Vert^2) \mathbf{d}^n.
\end{multline}

The first-order equations \eqref{LHS} can be rewritten in a similar fashion as a vector equation involving $\mathbf{x}^{n-1}, \mathbf{x}^n$ and $\mathbf{a}^{n-1}$. However, as there is no need to solve these equations directly (they merely serve to determine the slack variable $\mathbf{d}^n$), we will not go in further detail.

\paragraph{Summary.} To solve the discrete equations of motion \eqref{projected_discrete2} and \eqref{LHS} in the case of a non-$\mathbb{S}^1$-invariant Hamiltonian, we proceed as follows:

\fbox{%
\begin{minipage}{\textwidth}
\begin{enumerate}
	\item Given initial conditions $(\varphi^{n-1}, \varphi^n) \in \mathbb{S}^3 \times \mathbb{S}^3$, compute the slack $\mathbf{d}^n$ from \eqref{LHS}.
	\item For this $\mathbf{d}^n$, find $\mathbf{a}^n$ from \eqref{nonlinear_a}.
	\item Once $\mathbf{a}^n$ is known, update $\varphi^n$ to find $\varphi^{n+1}$ using the Cayley map \eqref{cayley_update}.
\end{enumerate}
\end{minipage}}

The advantage of computing $\varphi^{n+1}$ indirectly via $\mathbf{a}^n$ is that \eqref{nonlinear_a} is a nonlinear equation defined on $\mathfrak{su}(2)^N$. As this is a vector space, a standard nonlinear solver can be used to solve \eqref{nonlinear_a}. While \cite{OwWe2000} developed an extension of Newton's method that preserves the Lie group structure, it is much more computationally involved.

\bibliography{hopf-biblio}

\providecommand{\arxiv}[1]{\texttt{arXiv:#1}}\providecommand{\href}[1]{\texttt%
{#1}}\def\polhk#1{\setbox0=\hbox{#1}{\ooalign{\hidewidth\lower1.5ex\hbox{`}\hi%
dewidth\crcr\unhbox0}}} \def\cprime{$'$}
\begin{thebibliography}{64}
\expandafter\ifx\csname natexlab\endcsname\relax\def\natexlab#1{#1}\fi

\bibitem[Aref(2007)]{Ar2007}
Aref, H. [2007], Point vortex dynamics: a classical mathematics playground,
  {\em J. Math. Phys.} \textbf{48}, 065401, 23.

\bibitem[Aref(2011)]{Ar2011}
Aref, H. [2011], Relative equilibria of point vortices and the fundamental
  theorem of algebra, {\em Proceedings of the Royal Society A: Mathematical,
  Physical and Engineering Science} \textbf{467}, 2168--2184.

\bibitem[Benettin and Giorgilli(1994)]{BeGi1994}
Benettin, G. and A.~Giorgilli [1994], On the {H}amiltonian interpolation of
  near to the identity symplectic mappings with application to symplectic
  integration algorithms, {\em J. Stat. Phys.} \textbf{74}, 1117--1143.

\bibitem[Birkhoff(1966)]{Bi1966}
Birkhoff, G.~D. [1966], {\em Dynamical systems}.
\newblock With an addendum by Jurgen Moser. American Mathematical Society
  Colloquium Publications, Vol. IX. American Mathematical Society, Providence,
  R.I.

\bibitem[Boatto and Koiller(2008)]{BoKo2008}
Boatto, S. and J.~Koiller [2008], {\em Vortices on closed surfaces}, Preprint
  (arXiv:0802.4313v1), 2008.

\bibitem[Bogomolov(1977)]{Bo1977}
Bogomolov, V.~A. [1977], Dynamics of vorticity at a sphere, {\em Fluid
  Dynamics} \textbf{12}, 863--870.
\newblock 10.1007/BF01090320.

\bibitem[Borisov et~al.(2004)Borisov, Mamaev, and Kilin]{BoMaKi2004}
Borisov, A.~V., I.~S. Mamaev, and A.~A. Kilin [2004], Absolute and relative
  choreographies in the problem of point vortices moving on a plane, {\em
  Regul. Chaotic Dyn.} \textbf{9}, 101--111.

\bibitem[Bou-Rabee and Marsden(2009)]{BoMa2009}
Bou-Rabee, N. and J.~E. Marsden [2009], Hamilton-{P}ontryagin integrators on
  {L}ie groups. {I}. {I}ntroduction and structure-preserving properties, {\em
  Found. Comput. Math.} \textbf{9}, 197--219.

\bibitem[Boyland et~al.(2003)Boyland, Stremler, and Aref]{BoStAr2003}
Boyland, P., M.~Stremler, and H.~Aref [2003], Topological fluid mechanics of
  point vortex motions, {\em Physica D: Nonlinear Phenomena} \textbf{175}, 69
  -- 95.

\bibitem[Brown(2006)]{Brown2006}
Brown, J.~D. [2006], Midpoint rule as a variational-symplectic integrator:
  Hamiltonian systems, {\em Phys. Rev. D} \textbf{73}, 024001.

\bibitem[Bruveris et~al.(2011)Bruveris, Ellis, Holm, and
  Gay-Balmaz]{BrElHoGa2011}
Bruveris, M., D.~C.~P. Ellis, D.~D. Holm, and F.~Gay-Balmaz [2011],
  Un-reduction, {\em J. Geom. Mech.} \textbf{3}, 363--387.

\bibitem[Cendra and Marsden(1987)]{CeMa1987}
Cendra, H. and J.~E. Marsden [1987], Lin constraints, {C}lebsch potentials and
  variational principles, {\em Phys. D} \textbf{27}, 63--89.

\bibitem[Chamoun et~al.(2009)Chamoun, Kanso, and Newton]{ChKaNe2009}
Chamoun, G., E.~Kanso, and P.~K. Newton [2009], Von K{\'a}rm{\'a}n vortex
  streets on the sphere, {\em Physics of Fluids} \textbf{21}, 116603.

\bibitem[Chapman(1978)]{Ch1978}
Chapman, D. M.~F. [1978], Ideal vortex motion in two dimensions: Symmetries and
  conservation laws, {\em Journal of Mathematical Physics} \textbf{19},
  1988--1992.

\bibitem[Chartier et~al.(2010)Chartier, Darrigrand, and Faou]{ChDaFa2010}
Chartier, P., E.~Darrigrand, and E.~Faou [2010], A regular fast multipole
  method for geometric numerical integrations of {H}amiltonian systems, {\em
  BIT} \textbf{50}, 23--40.

\bibitem[Chorin(1973)]{Ch1973}
Chorin, A.~J. [1973], Numerical study of slightly viscous flow, {\em J. Fluid
  Mech.} \textbf{57}, 785--796.

\bibitem[Cotter and Holm(2009)]{CoHo2009}
Cotter, C.~J. and D.~D. Holm [2009], Continuous and discrete {C}lebsch
  variational principles, {\em Found. Comput. Math.} \textbf{9}, 221--242.

\bibitem[Eng{\o} and Faltinsen(2002)]{EnFa2002}
Eng{\o}, K. and S.~Faltinsen [2002], Numerical {I}ntegration of {L}ie-{P}oisson
  {S}ystems while {P}reserving {C}oadjoint {O}rbits and {E}nergy, {\em SIAM
  Journal on Numerical Analysis} \textbf{39}, pp. 128--145.

\bibitem[Faddeev and Jackiw(1988)]{FaJa1988}
Faddeev, L. and R.~Jackiw [1988], Hamiltonian reduction of unconstrained and
  constrained systems, {\em Phys. Rev. Lett.} \textbf{60}, 1692--1694.

\bibitem[Frankel(2004)]{Fr2004}
Frankel, T. [2004], {\em The {G}eometry of {P}hysics: {A}n {I}ntroduction}.
\newblock Cambridge University Press, Cambridge, second edition.

\bibitem[Gotay(1979)]{Go1979}
Gotay, M. [1979], {\em Presymplectic {M}anifolds, {G}eometric {C}onstraint
  {T}heory and the {D}irac-{B}ergmann {T}heory of {C}onstraints}, PhD thesis,
  University of Maryland.

\bibitem[Hairer(1994)]{Ha1994}
Hairer, E. [1994], Backward analysis of numerical integrators and symplectic
  methods, {\em Ann. Numer. Math.} \textbf{1}, 107--132.

\bibitem[Hairer and Lubich(1997)]{HaLu1997}
Hairer, E. and C.~Lubich [1997], The life-span of backward error analysis for
  numerical integrators, {\em Numer. Math.} \textbf{76}, 441--462.

\bibitem[Hairer et~al.(2002)Hairer, Lubich, and Wanner]{HaLuWa2002}
Hairer, E., C.~Lubich, and G.~Wanner [2002], {\em Geometric numerical
  integration}, volume~31 of {\em Springer Series in Computational
  Mathematics}.
\newblock Springer-Verlag, Berlin, 1st edition.

\bibitem[Khesin(2012)]{Kh2012}
Khesin, B. [2012], Symplectic structures and dynamics on vortex membranes, {\em
  Mosc. Math. J.} \textbf{12}, 413--434, 461--462.

\bibitem[Kidambi and Newton(1998)]{KiNe1998}
Kidambi, R. and P.~K. Newton [1998], Motion of three point vortices on a
  sphere, {\em Phys. D} \textbf{116}, 143--175.

\bibitem[Kimura and Okamoto(1987)]{KiOk1987}
Kimura, Y. and H.~Okamoto [1987], Vortex {M}otion on a {S}phere, {\em J. Phys.
  Soc. Japan} \textbf{56}, 4203--4206.

\bibitem[Kobilarov and Marsden(2011)]{KoMa2011}
Kobilarov, M. and J.~Marsden [2011], Discrete Geometric Optimal Control on Lie
  Groups, {\em IEEE Transactions on Robotics} \textbf{27}, 641 --655.

\bibitem[Kostant(2005)]{Ko2005}
Kostant, B. [2005], Minimal coadjoint orbits and symplectic induction.
\newblock In {\em The breadth of symplectic and {P}oisson geometry}, volume 232
  of {\em Progr. Math.}, pages 391--422. Birkh\"auser Boston, Boston, MA.

\bibitem[Lamb(1945)]{La1945}
Lamb, H. [1945], {\em Hydrodynamics}.
\newblock Dover Publications.
\newblock Reprint of the 1932 Cambridge University Press edition.

\bibitem[Lee et~al.(2007)Lee, Leok, and McClamroch]{LeLeMc2007}
Lee, T., M.~Leok, and N.~H. McClamroch [2007], Lie group variational
  integrators for the full body problem, {\em Comput. Methods Appl. Mech.
  Engrg.} \textbf{196}, 2907--2924.

\bibitem[Lee et~al.(2009)Lee, Leok, and McClamroch]{LeLeMc2009}
Lee, T., M.~Leok, and N.~H. McClamroch [2009], Lagrangian mechanics and
  variational integrators on two-spheres, {\em Internat. J. Numer. Methods
  Engrg.} \textbf{79}, 1147--1174.

\bibitem[Lema{\^{\i}}tre(1948)]{Le1948}
Lema{\^{\i}}tre, G. [1948], Quaternions et espace elliptique, {\em Pont. Acad.
  Sci. Acta} \textbf{12}, 57--78.

\bibitem[Leok and Zhang(2011)]{LeZh2011}
Leok, M. and J.~Zhang [2011], Discrete Hamiltonian variational integrators,
  {\em IMA Journal of Numerical Analysis} \textbf{31}, 1497--1532.

\bibitem[Leyendecker et~al.(2008)Leyendecker, Marsden, and Ortiz]{LeMaOr2008}
Leyendecker, S., J.~E. Marsden, and M.~Ortiz [2008], Variational integrators
  for constrained dynamical systems, {\em ZAMM Z. Angew. Math. Mech.}
  \textbf{88}, 677--708.

\bibitem[Lim et~al.(2001)Lim, Montaldi, and Roberts]{LiMoRo2001}
Lim, C., J.~Montaldi, and M.~Roberts [2001], Relative equilibria of point
  vortices on the sphere, {\em Phys. D} \textbf{148}, 97--135.

\bibitem[Ma and Rowley(2010)]{MaRo2010}
Ma, Z. and C.~W. Rowley [2010], Lie-{P}oisson integrators: a {H}amiltonian,
  variational approach, {\em Internat. J. Numer. Methods Engrg.} \textbf{82},
  1609--1644.

\bibitem[Majda and Bertozzi(2002)]{MaBe2002}
Majda, A.~J. and A.~L. Bertozzi [2002], {\em Vorticity and incompressible
  flow}, volume~27 of {\em Cambridge Texts in Applied Mathematics}.
\newblock Cambridge University Press, Cambridge.

\bibitem[Marsden and West(2001)]{MaWe2001}
Marsden, J.~E. and M.~West [2001], Discrete mechanics and variational
  integrators, {\em Acta Numerica} \textbf{10}, 357--514.

\bibitem[McDuff and Salamon(1998)]{McSa1998}
McDuff, D. and D.~Salamon [1998], {\em Introduction to symplectic topology}.
\newblock Oxford Mathematical Monographs. The Clarendon Press Oxford University
  Press, New York, second edition.

\bibitem[Milne-Thomson(1968)]{Mi1968}
Milne-Thomson, L. [1968], {\em Theoretical hydrodynamics.}
\newblock London: MacMillan and Co. Ltd., fifth edition, revised and enlarged
  edition.

\bibitem[Montgomery(2002)]{Mo2002}
Montgomery, R. [2002], {\em A tour of subriemannian geometries, their geodesics
  and applications}, volume~91 of {\em Mathematical Surveys and Monographs}.
\newblock American Mathematical Society, Providence, RI.

\bibitem[Moser and Veselov(1991)]{MoVe1991}
Moser, J. and A.~Veselov [1991], Discrete versions of some classical integrable
  systems and factorization of matrix polynomials, {\em Comm. Math. Phys.}
  \textbf{139}, 217--243.

\bibitem[Newton(2001)]{Ne2001}
Newton, P.~K. [2001], {\em The {$N$}-vortex problem. {A}nalytical techniques},
  volume 145 of {\em Applied Mathematical Sciences}.
\newblock Springer-Verlag, New York.

\bibitem[Newton and Sakajo(2011)]{NeSa2011}
Newton, P.~K. and T.~Sakajo [2011], Point vortex equilibria and optimal
  packings of circles on a sphere, {\em Proceedings of the Royal Society A:
  Mathematical, Physical and Engineering Science} \textbf{467}, 1468--1490.

\bibitem[Novikov(1982)]{No1982}
Novikov, S.~P. [1982], The Hamiltonian formalism and a many-valued analogue of
  Morse theory, {\em Russian Mathematical Surveys} \textbf{37}, 1.

\bibitem[Oh(1997)]{Oh1997}
Oh, Y.-G. [1997], Symplectic topology as the geometry of action functional.
  {I}. {R}elative {F}loer theory on the cotangent bundle, {\em J. Differential
  Geom.} \textbf{46}, 499--577.

\bibitem[Oliphant(2007)]{Ol2007}
Oliphant, T.~E. [2007], Python for Scientific Computing, {\em Computing in
  Science and Engg.} \textbf{9}, 10--20.

\bibitem[Onsager(1949)]{On1949}
Onsager, L. [1949], Statistical hydrodynamics, {\em Nuovo Cimento (9)}
  \textbf{6}, 279--287.

\bibitem[Owren and Welfert(2000)]{OwWe2000}
Owren, B. and B.~Welfert [2000], The {N}ewton iteration on {L}ie groups, {\em
  BIT} \textbf{40}, 121--145.

\bibitem[Pekarsky and Marsden(1998)]{PeMa1998}
Pekarsky, S. and J.~E. Marsden [1998], Point vortices on a sphere: stability of
  relative equilibria, {\em J. Math. Phys.} \textbf{39}, 5894--5907.

\bibitem[Polvani and Dritschel(1993)]{PoDr1993}
Polvani, L.~M. and D.~G. Dritschel [1993], Wave and vortex dynamics on the
  surface of a sphere, {\em J. Fluid Mech.} \textbf{255}, 35--64.

\bibitem[Pullin and Saffman(1991)]{PuSa1991}
Pullin, D.~I. and P.~G. Saffman [1991], Long-Time Symplectic Integration: The
  Example of Four-Vortex Motion, {\em Proceedings: Mathematical and Physical
  Sciences} \textbf{432}, pp. 481--494.

\bibitem[Reich(1999)]{Re1999}
Reich, S. [1999], Backward error analysis for numerical integrators, {\em SIAM
  J. Numer. Anal.} \textbf{36}, 1549--1570.

\bibitem[Rowley and Marsden(2002)]{RoMa2002}
Rowley, C. and J.~Marsden [2002], Variational integrators for degenerate
  Lagrangians, with application to point vortices.
\newblock In {\em Proceedings of the 41st IEEE Conference on Decision and
  Control}, volume~2, pages 1521 -- 1527.

\bibitem[Saffman(1992)]{Sa1992}
Saffman, P.~G. [1992], {\em Vortex dynamics}.
\newblock Cambridge Monographs on Mechanics and Applied Mathematics. Cambridge
  University Press, New York.

\bibitem[Sakajo(2004)]{Sakajo2004}
Sakajo, T. [2004], Motion of a vortex sheet on a sphere with pole vortices,
  {\em Phys. Fluids} \textbf{16}, 717--727.

\bibitem[Sakajo(2008)]{Sa2008}
Sakajo, T. [2008], Non-self-similar, partial, and robust collapse of four point
  vortices on a sphere, {\em Phys. Rev. E} \textbf{78}, 016312.

\bibitem[Shashikanth(2012)]{Sh2012}
Shashikanth, B.~N. [2012], Vortex dynamics in {$\Bbb R^4$}, {\em J. Math.
  Phys.} \textbf{53}, 013103, 21.

\bibitem[Souli{\`e}re and Tokieda(2002)]{SoTo2002}
Souli{\`e}re, A. and T.~Tokieda [2002], Periodic motions of vortices on
  surfaces with symmetry, {\em Journal of Fluid Mechanics} \textbf{460},
  83--92.

\bibitem[Urbantke(2003)]{Ur2003}
Urbantke, H.~K. [2003], The {H}opf fibration --- seven times in physics, {\em
  J. Geom. Phys.} \textbf{46}, 125--150.

\bibitem[von Helmholtz(1858)]{He1858}
von Helmholtz, H. [1858], \"{U}ber {I}ntegrale der hydro-dynamischen
  {G}leichungen, welche den {W}irbelbewegungen entsprechen, {\em Journal
  f{\"u}r die reine und angewandte Mathematik (Crelles Journal)} \textbf{55},
  25--55.

\bibitem[Woodhouse(1992)]{Wo1992}
Woodhouse, N. M.~J. [1992], {\em Geometric quantization}.
\newblock Oxford Mathematical Monographs. The Clarendon Press Oxford University
  Press, New York, second edition.
\newblock Oxford Science Publications.

\bibitem[Zhong and Marsden(1988)]{ZhMa1988}
Zhong, G. and J.~E. Marsden [1988], Lie-{P}oisson {H}amilton-{J}acobi theory
  and {L}ie-{P}oisson integrators, {\em Phys. Lett. A} \textbf{133}, 134--139.

\end{thebibliography}
\bibliographystyle{marsden}

\end{document}